\documentclass[floatfix,preprint,showemail,preprintnumbers,amsfonts,pre]{revtex4}
\usepackage{graphicx}
\usepackage{bm}

\begin{document}

\title{Self-optimization, community stability, and fluctuations
in two individual-based models of biological coevolution
}

\author{Per Arne Rikvold}\email{rikvold@scs.fsu.edu} 

\affiliation{School of Computational Science,
	      Center for Materials Research and Technology,\\
	      National High Magnetic Field Laboratory, and Department of
	      Physics,\\
	      Florida State University, Tallahassee, Florida 32306-4120,
	      USA\\
	      and Department of Fundamental Sciences,
	      Faculty of Integrated Human Studies,
	      Kyoto University, Kyoto 606, Japan
	      \\
}

\date{\today}

\begin{abstract}
We compare and contrast the long-time dynamical properties of
two individual-based models of biological
coevolution. Selection occurs via multispecies, stochastic
population dynamics with reproduction probabilities that 
depend nonlinearly on the population densities of all species
resident in the community. New species are introduced through 
mutation. Both models are amenable to exact
linear stability analysis, and we compare the analytic results
with large-scale kinetic Monte Carlo simulations, obtaining the 
population size as a function of an average interspecies
interaction strength. Over time,
the models {\it self-optimize\/} through mutation and
selection to approximately maximize a community 
fitness
function, subject only to constraints internal to the particular model.
If the interspecies
interactions are randomly distributed on an interval including
positive values, the system evolves toward self-sustaining, 
{\it mutualistic\/} communities.
In contrast, for the predator-prey case the matrix of interactions is
antisymmetric, and a nonzero population size must be sustained by an
external resource.
Time series of the diversity and population size for both 
models show approximate $1/f$ noise and power-law distributions for the
lifetimes of communities and species.
For the mutualistic model, these two lifetime distributions have the
same exponent, while their exponents
are different for the predator-prey model.
The difference is probably due to greater resilience
toward mass extinctions in the
food-web like communities produced by the predator-prey model.
\end{abstract}

\maketitle


\section{Introduction}
\label{sec:Int}

Traditionally, problems in ecology and evolution have been addressed at
very different levels of resolution. Typically, ecological problems are
studied on a timescale of generations and often at the level of
individual organisms, while issues in evolution are considered on much
longer, often geological, timescales and usually at the level of
species or higher-level taxa. However, in recent years it has been
recognized that processes at the ecological and evolutionary scales can
be strongly linked \cite{DROS01B,THOM98,THOM99,YOSH03}. 
Several models have therefore been proposed, which aim to
model the complex problem of coevolution in a fitness
landscape that changes with the composition of the community, 
while spanning the disparate scales of both temporal and taxonomic resolution. 
Early steps in this direction were simulations of parapatric and 
sympatric speciation \cite{CROS70} and the coupled $NK$ model with
population dynamics \cite{KAUF93,KAUF91}.  
More recent contributions include the Webworld model
\cite{CALD98,DROS01B,DROS04},
the Tangled-nature model
\cite{CHRI02,COLL03,HALL02}
and simplified versions of the latter
\cite{RIKV03,SEVI06,ZIA04}, 
as well as network models  
\cite{CHOW05,CHOW03A}.
Recently, large individual-based simulations have also been performed of
parapatric and sympatric speciation \cite{GAVR98,GAVR00} and of
adaptive radiation \cite{GAVR05}. 
Many of these models are deliberately quite simple, aiming to
elucidate {\it universal\/} features that are largely independent of the
finer details of the ecological interactions and the evolutionary
mechanisms. Such universal 
features may include lifetime distributions for species and communities,
as well as other aspects of extinction statistics,
statistical properties of fluctuations in diversity and population 
sizes, and
the structure and dynamics of food webs that develop and change 
with time. By changing
specific features of such simplified models, one hopes to learn
which aspects influence the observed properties of the
resulting communities and their development with time. 

In this paper we compare and contrast two stochastic
coevolution models that combine
features from some of those mentioned above. 
In each, selection is provided by an individual-based
population dynamics, while new species are provided by a low rate of
mutation. The models are studied 
both analytically by linear stability theory, and numerically 
by large-scale kinetic Monte Carlo simulations. 
The first model (Model A) allows direct mutualistic 
interspecies interactions, and some of its properties were discussed 
previously \cite{RIKV03,SEVI06,ZIA04}. 
It is a simplified version of the Tangled-nature model of Jensen and
coworkers \cite{CHRI02,COLL03,HALL02}. 
The second model (Model B) is a predator-prey model. For both models we 
obtain exactly the fixed-point mean population sizes and stability
properties for any given community of species in the limit of vanishing 
mutation rate, using linear stability theory. These results
enable us to define a community 
fitness 
function, cubic in the 
mean total population size, that is maximized at the fixed point 
for a given community of species. The fact that this much information
can be obtained analytically makes these models 
uniquely well suited as benchmarks for
more elaborate, but less tractable, nonlinear models.  

The analytical results are followed by numerical simulations of the
dynamics of both models for nonzero mutation rates. We focus
on very long simulations in a regime where both diversity and
population size are statistically stationary, albeit with very large, 
strongly correlated fluctuations and an intermittently
vigorous turnover of species. We thus 
study the {\it intrinsic\/} dynamics of extinction and origination of
species and communities in the absence of external perturbations.
Understanding of these intrinsic fluctuations in the stationary state
should enable one to estimate the community's response to external
perturbations \cite{SATO03} in a way analogous to the fluctuation-dissipation
relations of statistical mechanics \cite{PATH96}. 

Three main conclusions emerge from this combined analytical and numerical
investigation. 
\begin{itemize}
\item
Mutations enable the models to evolve communities that tend to
maximize the community 
fitness 
function, subject only to
constraints that are internal to the specific model. Although there is
vigorous turnover of species and communities, all communities that
persist for a significant time remain in the vicinity of this maximum. 
\item
The simulated systems exhibit power-law distributions in the lifetimes of 
individual species, as well as of communities, and the fluctuations 
in diversity and population size are characterized by approximate $1/f$ noise.
\item
The two models show distinct differences in the community turnover,
which are reflected in differences between the power-law exponent of the
community-lifetime distribution. In Model A, extinctions of species tend
to be highly synchronized, such that a whole community collapses in a mass
extinction on a relatively short time scale. In Model B, on the other hand, 
extinctions are more often limited to a subset of the resident species. 
This can be explained by differences between
the structures of the interaction networks characterizing 
long-lived communities in the two models. For Model B, this network
amounts to a simple food web. 
\end{itemize}

The rest of this paper is organized as follows. 
The models are introduced in Sec.~\ref{sec:Mod}. 
In Sec.~\ref{sec:Lin} we perform a full analytical linear stability analysis, 
and the results are compared with large-scale kinetic Monte Carlo simulations. 
The simulations are described in further detail in 
Sec.~\ref{sec:Dyn}, where lifetime distributions and power spectral
densities also are reported. A concluding summary is presented in 
Sec.~\ref{sec:Conc}. Some technical details of the derivation of the
community 
fitness 
function are given in Appendix~\ref{sec:AppA}, 
an analytical result for the optimum population size of Model A is
obtained in Appendix~\ref{sec:AppX}, and
a discussion of the consequences of changes in several of the model
parameters is found in Appendix~\ref{sec:AppB}.

\section{The Models}
\label{sec:Mod}

In both models, selection is provided by the reproduction rates in 
an individual-based, simplified, stochastic 
population-dynamics model with nonoverlapping generations. 
This interacting birth-death process is  
augmented to enable evolution of new species by a mutation mechanism. 
The mutations act on a haploid, binary ``genome" of length $L$, as 
introduced in Eigen's model for molecular evolution 
\cite{EIGE71,EIGE88}. This bit string defines the species, which are 
identified by the integer label $I \in [0,2^L-1]$. 
Typically, only a few of these $2^L$ potential species 
are resident in the community at any one time.  

Individual organisms of species $I$ reproduce asexually at the 
end of each generation, each giving rise to $F$ offspring individuals
with probability $P_I$ before dying. 
With probability $(1-P_I)$, they die without offspring. No individual 
thus survives beyond one generation. 
(For simplicity, the fecundity $F$ is assumed fixed, independent of both
species and individual.) 

During reproduction, each gene in an offspring individual's genome
may undergo mutation ($0 \rightarrow 1$ or $1 \rightarrow 0$) 
with a small probability, $\mu/L$.
The mutation thus corresponds to diffusional moves from corner to corner
along the edges of an $L$-dimensional hypercube \cite{GAVR99,GAVR04}. 
A mutated individual
is assumed to belong to a different species than its parent, with 
different properties. Genotype and phenotype are thus in one-to-one 
correspondence in these models. 
This is clearly a highly idealized picture, and it is introduced 
to maximize the pool of different species available within the computational 
resources. The approximation is justified by
a large-scale computational study of 
a version of Model A, in which species that differ by as many as 
$L/2$ bits have correlated properties \cite{SEVI06}. 
Quite remarkably, this study reveals that the correlated model has  
long-time dynamical properties very similar to 
the uncorrelated Model A studied here.

The reproduction probability $P_I(t)$ for an individual of species $I$  
in generation $t$ depends on the individual's ability to  
utilize the amount of external resources available, $R$, 
and on its interactions with the population sizes $n_J(t)$ 
of all the species present in the community at that time. 
The dependence of $P_I$ on the set of $n_J$ is determined by an {\it
interaction matrix\/} $\bf M$ \cite{SOLE96B} 
with elements ${M}_{IJ} \in [-1,1]$ in a
way defined specifically in the next paragraph. 
We emphasize that $\bf M$ is chosen randomly at the beginning of each
simulation run and is subsequently kept {\it constant\/} throughout the run. 
If ${M}_{IJ}$ is positive and
${M}_{JI}$ is negative, then $I$ is a predator and $J$ its prey, and {\it vice
versa\/}. If both matrix elements are positive, the species interact 
directly in a mutualistic way, while both elements
negative implies direct competition. 

Specifically, the reproduction probability for species $I$, $P_I(t)$,
depends on $R$ and the set $\{n_J(t)\}$ through the nonlinear form,
\begin{equation}
P_I(t) = \frac{1}{1 + \exp[-\Delta_I(R,\{n_J(t)\})]} \;,
\label{eq:PI}
\end{equation}
with the density dependent argument 
\begin{equation}
\Delta_I(R,\{n_J(t)\}) = - b_I + \eta_I R / N_{\rm tot}(t)
+ \sum_J M_{IJ} n_J(t) / N_{\rm tot}(t) - N_{\rm tot}(t)/N_0
\;.
\label{eq:Delta}
\end{equation}
Here $b_I$ can be seen as the ``cost" of reproduction (always
positive), and $\eta_I$ (positive for primary producers or autotrophs, 
and zero for consumers or heterotrophs) is the ability of individuals of
species $I$ to utilize the external resource $R$. The latter 
is renewed at the same level each generation. 
It does {\it not\/} have independent dynamics.
The total population size is $N_{\rm tot}(t) = \sum_J n_J(t)$, 
and the constant $N_0$ is an environmental carrying capacity 
\cite{MURR89,VERH1838} 
that prevents $N_{\rm tot}(t)$ from diverging to infinity. 
It may be seen as representing implicit resource limitations not
explicitly included in $R$, such as available space. 

For large positive $\Delta_I$, 
the individual almost certainly reproduces, giving rise to $F$
offspring, while for large negative $\Delta_I$, 
it almost certainly dies without offspring. 
The reproduction probability $P_I$, together with its argument
$\Delta_I$, play the role of a functional response for 
this class of models \cite{DROS01B,KREB01}. 
The two externally determined parameters that influence the population
size, the resource term $R$ and the carrying capacity $N_0$, play very
different roles. From Eq.~(\ref{eq:Delta}) it is seen that $R$ encourages
population growth, especially for small total population sizes, while it
has very little effect for large $N_{\rm tot}$. It can thus be thought
of as representing a finite amount of available food. The carrying
capacity $N_0$, on the other hand, always discourages growth, but its
effect is only appreciable for large population sizes. It cannot maintain
a nonzero population by itself and is thus best thought of as
representing an overall limitation, such as finite available space. 
If $R=0$, a nonzero population can 
therefore 
only be maintained by
mutually positive interspecies interactions. This rather unrealistic
aspect is shared by the Tangled-nature model \cite{CHRI02,COLL03,HALL02}. 

In this work, 
the model parameters are chosen to represent the realistic situation that 
the number of species resident in the community at any time
is much smaller than the number of potential species 
(i.e., that $\mathcal{N}(t) \ll 2^L$), and also that 
$\mathcal{N}(t) \ll N_{\rm tot}(t)$.  

An analytic approximation describing the time development of the mean
population sizes (averaged over independent realizations), 
$\langle n_I(t) \rangle$, can be
written as a set of coupled difference equations,
\begin{eqnarray}
\langle n_I(t+1) \rangle
&=& \langle n_I(t) \rangle FP_I(R,\{\langle n_J(t)\rangle \})[1-\mu ] 
\nonumber\\ 
&& +(\mu/L)F\sum_{K(I)}\langle n_{K(I)}(t)
\rangle P_{K(I)}(R,\{\langle n_J(t) \rangle \}) + O(\mu^2)
\;,
\label{eq:MF}
\end{eqnarray}
where $K(I)$ is the set of species that can be generated from species $I$
by a single mutation (``nearest neighbors" of $I$ in genotype space).

\subsection{Model A}
\label{sec:modA}
Model A was introduced and studied in \cite{RIKV03,ZIA04}. 
In this model, the $M_{IJ}$ for $I \neq J$ are
stochastically independent and uniformly distributed on $[-1,+1]$, 
while the intra-species interactions
$M_{II} = 0$. The external resource $R$ and the reproduction
costs $b_I$ are equal to zero, and the total population size $N_{\rm
tot}(t)$ is limited only by the carrying capacity $N_0$.
The model is found to evolve through a succession of quasi-stable,
mutualistic communities. 

\subsection{Model B}
\label{sec:modB}
Model B is a predator-prey model. 
This is implemented by making the off-diagonal 
part of $\bf M$ antisymmetric. In order to keep the connectance of
the resulting communities consistent with food webs
observed in nature \cite{DUNN02,GARL04}, 
the $(M_{IJ},M_{JI})$ pairs are
chosen nonzero with probability $c = 0.1$. The nonzero 
elements in the upper triangle of $\bf M$ are
chosen independently and uniformly on $[-1,+1]$. 
This model does not include a population-limiting carrying capacity
[i.e., formally, $N_0 = \infty$ in Eq.~(\ref{eq:Delta})], 
and the community is supported by a constant external resource, $R$. 
Only a proportion $p$ of the $2^L$ potential species are producers 
that can directly utilize the resource 
(for the numerical data reported here, we use $p = 0.05$). 
Thus, with probability $(1-p)$ the resource coupling $\eta_I = 0$, 
representing consumers, while with probability $p$ the $\eta_I$
are independently and uniformly distributed on $(0,+1]$, representing 
producers of varying efficiency. 
In addition to these constraints on $\bf M$, we require 
that producers ($\eta_I>0$) always are the prey of consumers ($\eta_I=0$). 
In other words: the case $\eta_I>0$ and $\eta_J=0$ with $M_{IJ} \equiv -
M_{JI} >0$ is forbidden. 
Whenever it occurs, this situation is corrected during the
construction of $\bf M$ by reversing the signs of $M_{IJ}$ and
$M_{JI}$ for the pair in question. 
Consumers at higher trophic levels are allowed, and they are indeed 
observed in the resulting communities (see further discussion in
Sec.~\ref{sec:Dyn}). 
The population sizes are limited by
independent reproduction costs $b_I$ that are uniformly distributed on
$(0,+1]$, and by negative intra-species interactions $M_{II}$ 
independently and uniformly distributed on $[-1,0)$. 
The model is found to evolve through a succession of quasi-stable,
food-web like communities. 
Some preliminary numerical results were presented  in \cite{RIKV05A}.

\section{Linear Stability Analysis}
\label{sec:Lin}

The form of $\Delta_I$ in Eq.~(\ref{eq:Delta})  
represents frequency-dependent interactions that describe
universal competition and absence of adaptive foraging, and so is not 
very realistic. However, the normalization by $N_{\rm tot}(t)$  
has the advantage that it turns Eq.~(\ref{eq:MF})  with $\mu = 0$ 
into a set of linear equations for the average
population sizes in the fixed-point
community corresponding to a particular set of species. 
The set of equations can be solved analytically to give
the average total population size, as well as 
the average population size of each species
and the stability properties of the community. 
This mean-field level analysis also enables us to construct a community
fitness 
function that for a given set of species
is maximized by the fixed-point community, and
whose width is proportional to the size of the population fluctuations
around the fixed point, caused by the statistical birth-death process. 
(For a detailed discussion of the fluctuations in Model A in the absence
of mutations, see \cite{ZIA04}.) 
For small mutation rates, this picture remains valid as a description of
the population fluctuations on relatively short time scales, where the
species composition of the community remains constant, except for small
populations of unsuccessful mutants. 

\subsection{Fixed-point communities}
\label{sec:fpc}

\begin{figure}[tbp]
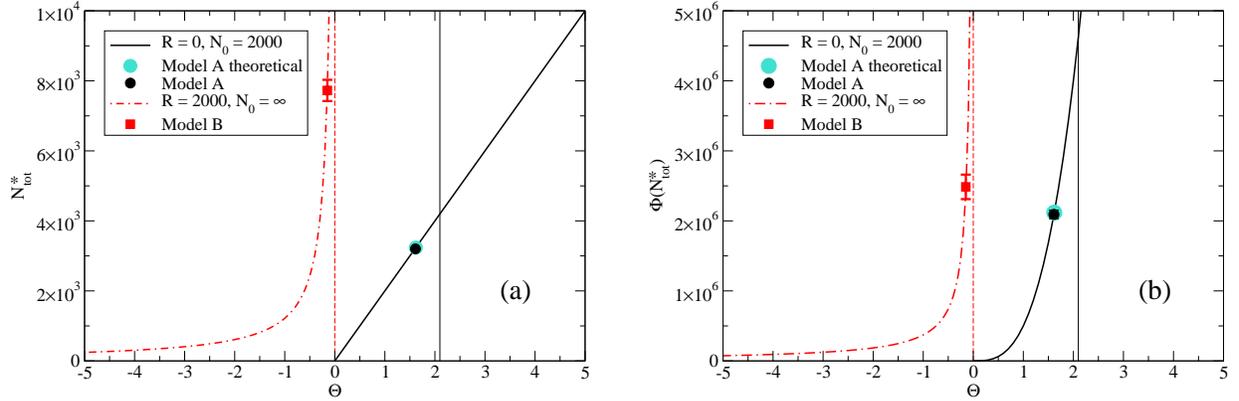
 
\begin{center}
\vspace*{0.3truecm}
\includegraphics[angle=0,width=.47\textwidth]{Phasefiga2.eps} 
\hspace{0.5truecm}
\includegraphics[angle=0,width=.47\textwidth]{Phasefigb2.eps} 
\end{center}
\caption[]{
(Color online.) 
Solutions for {\bf (a)} 
the total fixed-point population size $N_{\rm tot}^*$ 
and {\bf (b)} the maximum community 
fitness 
function 
$\Phi_{\rm max} = \Phi(N_{\rm tot}^*)$.  Both are shown vs $\Theta$ for fixed 
$\mathcal{E} = 0.61$.  The solid vertical lines represent the absolute
maximum value for $\Theta$ in Model A, 
while the dotted vertical lines represent the analogous limit for
Model B. The points marked ``Model A" and ``Model B" were obtained
from large-scale Monte Carlo simulations as described in the text.  
The calculation of the point marked ``Model A theoretical" is described in
Appendix~\protect\ref{sec:AppX}. 
}
\label{fig:ntot}
\end{figure}

To obtain the stationary solution of Eq.~(\ref{eq:MF}) with $\mu = 0$ 
for a community of $\mathcal{N}$ species, we require $P_I=1/F$ 
for all $\mathcal{N}$ species. 
Equations (\ref{eq:PI}) and~(\ref{eq:Delta}) then yield the
$\mathcal{N}$ linear relations 
\begin{equation}
- \tilde{b}_I + \eta_I \frac{R}{N_{\rm tot}^*} 
+ \sum_J M_{IJ} \frac{n_J^*}{N_{\rm tot}^*} 
- \frac{N_{\rm tot}^*}{N_0} =0
\;,
\label{eq:ss}
\end{equation}
where $\tilde{b}_I = b_I - \ln(F-1)$. (For simplicity, we 
have dropped the $\langle \; \rangle$ notation for the average
population sizes. The asterisk superscripts denote fixed-point
solutions.) In a vector notation where $\vec{v}$ is a column 
vector and $\vec{v}^T$ its conjugate row vector, 
$\vec{n}^*$ is the column
vector composed of the $\mathcal{N}$ nonzero $n_I^*$, while 
$\vec{1}^T$ is an $\mathcal{N}$-dimensional
row vector composed entirely of ones. 
Thus, the total population size is given by 
$N_{\rm tot}^* \equiv \sum_I n_I^* = \vec{1}^T\vec{n}^*$, and
Eq.~(\ref{eq:ss}) takes the matrix form
\begin{equation}
- \vec{\tilde{b}} N_{\rm tot}^* 
+ \vec{\eta} R  
+ {\hat{{\bf M}}} \vec{n}^* 
- \vec{1} (N_{\rm tot}^*)^2/N_0 = 0
\;.
\label{eq:ss2}
\end{equation}
Here, $\vec{\tilde{b}}$ is the column vector whose elements are
$\tilde{b}_I$, 
$\vec{\eta}$ is the column vector whose elements are $\eta_I$ 
(in both cases including only those $\mathcal{N}$
species that have nonzero $n_I^*$), 
$\hat{\bf M}$ is the corresponding $\mathcal{N}
\times \mathcal{N}$ submatrix of $\bf M$, and $\vec{1}$ is an 
$\mathcal{N}$-dimensional column vector of ones. 
The solution for $\vec{n}^*$ is 
\begin{equation}
\vec{n}^* = - {\hat{\bf M}}^{-1} \left[ \vec{\eta} R 
- \vec{\tilde{b}} N_{\rm tot}^*
- \vec{1} (N_{\rm tot}^*)^2 /N_0 \right]
\;,
\label{eq:ssn}
\end{equation}
where ${\hat{\bf M}}^{-1}$ is the inverse of ${\hat{\bf M}}$. 
(See below for a discussion of the effects of a singular $\hat{\bf M}$.) 
To find each $n_I^*$, we must first obtain 
$N_{\rm tot}^* \equiv \vec{1}^T \vec{n}^*$.
Multiplying Eq.~(\ref{eq:ssn}) from the left by $\vec{1}^T$, we
obtain the quadratic equation for $N_{\rm tot}$,  
\begin{equation}
R \mathcal{E} + \Theta N_{\rm tot}^* - (N_{\rm tot}^*)^2/N_0 = 0
\;.
\label{eq:quad}
\end{equation}
The coefficients, 
\begin{eqnarray}
\Theta =  
\frac{1 - \vec{1}^T {\hat{\bf M}}^{-1}  \vec{\tilde{b}}}
     {\vec{1}^T  {\hat{\bf M}}^{-1} \vec{1} }
& \;\; \mbox{\rm and} \;\; &
\mathcal{E} = \frac{\vec{1}^T {\hat{\bf M}}^{-1} \vec{\eta}}
                   {\vec{1}^T {\hat{\bf M}}^{-1} \vec{1} }
\;,
\label{eq:quad2}
\end{eqnarray}
can be viewed as an effective interaction strength and an effective 
coupling to the external resource, respectively. Approximate expressions 
for $\Theta$ and $\mathcal{E}$ that are less accurate but 
more intuitive are obtained in Appendix~\ref{sec:AppA}. 
The nonnegative solution of Eq.~(\ref{eq:quad}) is 
\begin{equation}
N_{\rm tot}^* = \frac{\Theta N_0}{2}
+ \sqrt{\left(\frac{\Theta N_0}{2} \right)^2 
+ R \mathcal{E} N_0 } 
\;.
\label{eq:sol}
\end{equation}
Figure~\ref{fig:ntot}(a) shows $N_{\rm tot}^*$ as a function of 
$\Theta$ for two choices of $N_0$ and $R$ at fixed $\mathcal{E}$. 
Special cases of the solution are
\begin{equation}
N_{\rm tot}^* = 
\left\{
\begin{array}{lllll}
\Theta {N_0}
& \mbox{for} & R=0 &\mbox{and}& \Theta \ge 0 \nonumber\\
0 
& \mbox{for} & R=0 &\mbox{and}& \Theta \le 0 \nonumber\\
\sqrt{R \mathcal{E} N_0 }
& \mbox{for} & \Theta = 0 
& \mbox{and} & \mathcal{E} \ge 0
\nonumber\\
- R \mathcal{E} / \Theta 
& \mbox{for} & N_0 = \infty & \mbox{and/or}& 
\vec{1}^T \hat{\bf M}^{-1} \vec{1} = 0 
\end{array}
\right.
\;.
\label{eq:spec}
\end{equation}
To find each $n_I^*$ separately, we now only need to insert 
$N_{\rm tot}^*$ in Eq.~(\ref{eq:ssn}). 

Only those $\vec{n}^*$ that have all positive elements can represent a 
{\it feasible\/} community \cite{ROBE74}. 
If ${\hat{\bf M}} = {\bf 0}$ or is otherwise singular, 
the set of equations (\ref{eq:ss2}) is 
inconsistent for $\mathcal{N} >1$, unless $\tilde{b}_I$ and $\eta_I$
both are independent of $I$.
The only possible stationary community then consists of one
single species, the one with the largest value of $-\tilde{b}_I$ for 
Model A or the one with the largest $\eta_I/\tilde{b}_I$ for Model B. 
This is a trivial example of competitive
exclusion \cite{ARMS80,DENB86,HARD60}, and stable multispecies 
coexistence in these models requires a nonsingular interaction 
matrix \footnote{Neutral versions of these models can be constructed by 
setting ${\hat{\bf M}} = {\bf 0}$, and then
requiring $-\tilde{b}_I$ or $\eta_I/\tilde{b}_I$ to be independent
of $I$ for Model A or Model B, respectively. This makes all species 
equivalent, and these neutral models thus
maintain biodiversity through genetic drift in the sense of Hubbell's neutral 
model \cite{ALON04,HUBB01,VOLK05,WILL06}.}. 

\begin{figure}[tbp] 
\begin{center}
\vspace*{0.3truecm}
\includegraphics[angle=0,width=.47\textwidth]{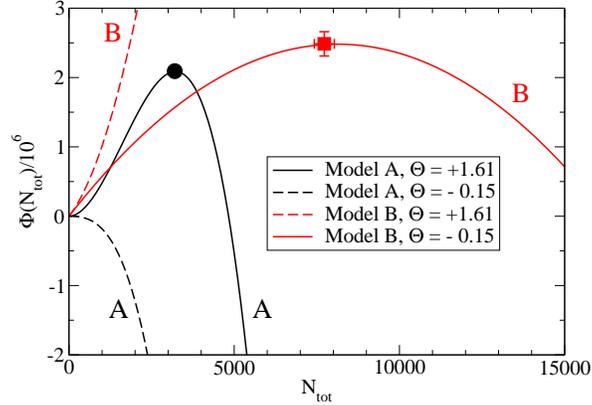} 
\end{center}
\caption[]{
(Color online.)
The community 
fitness 
function $\Phi$, shown vs $N_{\rm
tot}$ for Model A (black) and Model B (gray, red online). 
For Model A the parameters are: $F=4$, $R=0$, and $N_0=2000$. 
For Model B they are: $F=2$, $R=2000$, $\mathcal{E} = 0.61$,
and $N_0=\infty$. The values of $\Theta$ used are $1.61$, corresponding
to the average value taken by Model A in the simulations, 
and $-0.15$, corresponding to the average value taken by Model B. 
$\Phi(N_{\rm tot})$ has a nontrivial maximum for Model A at positive
$\Theta$ (solid black curve marked A), and for Model B at negative $\Theta$
(solid gray curve marked B (red online)). 
The circular and square data points are the results of Monte Carlo
simulations and correspond to the
equally shaped points in Fig.~\ref{fig:ntot}. They lie very close
to the maximum of $\Phi(N_{\rm tot})$ for each model. 
The dashed curves show the physically inaccessible cases of Model A with
$\Theta < 0$ (stable, absorbing state at $N_{\rm tot} = 0$) and Model B with 
$\Theta > 0$ (monotonically increasing $\Phi(N_{\rm tot})$). 
}
\label{fig:phi3}
\end{figure}
Equation~(\ref{eq:quad}) can be seen as a maximization condition
for a community 
fitness 
function, 
\begin{equation}
\Phi(N_{\rm tot}) 
= \left( 1 - \frac{1}{F} \right) 
\left(
R \mathcal{E} N_{\rm tot} 
+ \frac{\Theta}{2} N_{\rm tot}^2
- \frac{1}{3 N_0}N_{\rm tot}^3
\right)
\;,
\label{eq:fit}
\end{equation}
which is cubic in $N_{\rm tot}$ \footnote{Here we have 
obtained Eq.~(\protect\ref{eq:fit}) simply by integration of
Eq.~(\protect\ref{eq:quad}). 
A derivation that also
explains the prefactor $(1 - 1/F)$ and provides intuitive approximations
for $\Theta$ and $\mathcal{E}$ is given in
Appendix~\protect\ref{sec:AppA}.}. 
The dependence of $\Phi$ on $N_{\rm tot}$ is shown in
Fig.~\ref{fig:phi3} for Models A and B at two different values of
$\Theta$. In each model, the maximum of $\Phi$ represents the
fixed-point value of $N_{\rm tot}$, while its width gives a measure of
the extent of the fluctuations about the fixed-point population in a
community of fixed composition \cite{ZIA04}. 
The maximum value of $\Phi$ with respect to $N_{\rm tot}$, 
$\Phi_{\rm max} = \Phi(N_{\rm tot}^*)$, is shown in Fig.~\ref{fig:ntot}(b) 
vs $\Theta$ for two different values of $N_0$ at fixed $\mathcal{E}$. 

From Eq.~(\ref{eq:spec}) and Fig.~\ref{fig:ntot}(a) it is seen that  
Model A (finite $N_0$ and $R=0$) undergoes a transcritical 
bifurcation or exchange of stabilities \cite{CRAW91,STRO94}
at $\Theta = 0$. For $\Theta \le 0$, $N_{\rm tot}$  vanishes, while 
for $\Theta > 0$ it increases linearly with $\Theta$ \footnote{In the presence 
of a positive external resource ($R > 0$) the bifurcation for Model A 
would become {\it imperfect\/} 
\protect\cite{STRO94}, yielding positive 
$N_{\rm tot}$ for all $\Theta$, as is easily seen from 
Eq.~(\protect\ref{eq:sol}).}. Such behavior as a 
function of a control parameter is common in many nonequilibrium 
systems. In addition to population dynamics and the logistic map 
(to which the present models with $\mu = 0$ are closely related), 
these also include lasers and autocatalytic chemical reactions 
\cite{HAKE77,STRO94}.
On the other hand, for Model B ($N_0 = \infty$ and $R>0$), 
$N_{\rm tot}$ diverges to infinity as $\Theta$ approaches zero from below, 
and it would be infinite for $\Theta \ge 0$. 

The results discussed in the previous paragraphs would not be 
of evolutionary relevance 
if $\Theta$ were just an externally fixed control parameter, as it would 
be in the absence of extinctions and mutations. 
However, mutations enable the community not only to
maximize $\Phi(N_{\rm tot})$ for fixed parameters, {\it but also
to change the parameters in response to new mutations, leading to the
possibility of further increasing $\Phi_{\rm max}$\/}.
Numerical results from the large-scale kinetic Monte Carlo simulations
discussed in Sec.~\ref{sec:Dyn} show that, on average, 
the communities initially progress toward, 
and later settle down to fluctuate near, the
maximum value of $\Theta$ (and thus the maximum value of $\Phi_{\rm max}$) 
compatible with the constraints on $\bf M$ and $\tilde{\vec b}$. 
These constraints depend on the specific model as follows. 
For Model A, which allows direct mutualistic interactions 
(i.e., $M_{IJ}$ and $M_{JI}$ both positive), $\Theta$ can be positive, thus 
enabling communities with nonzero population size, even for $R=0$. 
For the predator-prey 
Model B, on the other hand, the antisymmetric structure of $\bf M$ forces 
$\Theta$ to be nonpositive. 
A detailed discussion of the time evolution of the mean-field parameters, 
$\Theta$, $\mathcal{E}$, $N_{\rm tot}^*$, and $\Phi_{\rm max}$, is given 
for both models in Sec.~\ref{sec:Dyn1}. 

In Monte Carlo simulations of Model A (in which the off-diagonal 
elements of $\bf M$ are uncorrelated and
uniformly distributed on $[-1,+1]$), it was found that,
after the initial period characterized by an average positive trend in
the mean-field parameters, 
the community spent most of its time in a succession of
quasi-steady states (QSSs), separated by brief bursts of intense 
evolutionary activity \cite{RIKV03}. 
All the QSSs studied in detail were found to be
mutualistic, with $\overline{M_{IJ}} = 0.78\pm0.03$.
[Here, the overbar represents averages over all the ten QSSs listed in
Table~I of \cite{RIKV03}.] The average of $N_{\rm tot}$, taken
over all the 16 realizations of $2^{25}$ generations that
were studied, was ${\overline{N_{\rm tot}}} = 3201\pm8$. 
(The average over only the ten QSSs agrees with the total average to within the 
statistical errors, showing that the periods when the system is not 
in a QSS contribute negligibly to the overall time averages.)
From these averages we can use Eqs.~(\ref{eq:spec}) and~(\ref{eq:fit}) 
to estimate the average mean-field quantities, 
$\overline{\Theta} = 1.61\pm0.01$ and 
$\overline{\Phi_{\rm max}} = (2.09 \pm 0.06)\times 10^6$. 
In Fig.~\ref{fig:ntot}(a), ${\overline{N_{\rm tot}}}$ is shown vs 
$\overline{\Theta}$ as a black dot, while the
corresponding value of $\overline{\Phi_{\rm max}}$ is shown 
the same way in Fig.~\ref{fig:ntot}(b). 

The results for Model B that are included in Fig.~\ref{fig:ntot}
were also obtained from Monte Carlo simulations of 2$^{25}$ generations. 
(See Sec.~\ref{sec:Dyn} for further details.)
The parameter values used in the figure
were extracted as the average values from thirteen 
QSS communities identified at late times in twelve independent simulation runs: 
$\overline{\mathcal{E}} = 0.61\pm0.04$ 
and $\overline{\Theta} = - 0.15\pm0.01$. 
The value of the total population size 
given in the figure is the total time average
over the simulations, averaged over all twelve runs,  
${\overline{N_{\rm tot}}} = 7726 \pm 303$. 
(Like for Model A, averages over only the 
particular observed QSS communities 
agree with this total average to within the error bars.)
The large uncertainty in ${\overline{N_{\rm tot}}}$ 
is a direct consequence of the steep slope of $N_{\rm tot}^*$ versus 
$\Theta$ closely below $\Theta = 0$ 
for Model B (see Fig.~\ref{fig:ntot}(a)). 
From these estimates and Eq.~(\ref{eq:fit}) we further get the estimate 
$\overline{\Phi_{\rm max}} = (2.49 \pm 0.18)\times 10^6$.

\subsection{Stability of fixed-point communities}
\label{sec:stab}

The internal stability of an $\mathcal{N}$-species 
fixed-point community is obtained from the matrix of partial derivatives, 
\begin{equation}
\left. \frac{\partial n_I(t+1)}{\partial n_J(t)} \right|_{\vec{n}^*} 
= 
\delta_{IJ} + \Lambda_{IJ} \;,
\label{eq:stab}
\end{equation}
where $\delta_{IJ}$ is the Kronecker delta and 
$\Lambda_{IJ}$ are elements of the {\it community matrix\/} 
$\bf \Lambda$ \cite{MURR89}. Straightforward differentiation yields
\begin{equation}
\Lambda_{IJ} = \left( 1 - \frac{1}{F} \right) \frac{n_I^*}{N_{\rm tot}^*} 
\left[ M_{IJ} 
- \frac{R \eta_I +({\hat{\bf M}}\vec{n}^*)_I}{N_{\rm tot}^*}
- \frac{1}{N_0}
\right]
\;,
\label{eq:lam}
\end{equation}
where $({\hat{\bf M}}\vec{n}^*)_I$ is the element of 
${\hat{\bf M}}\vec{n}^*$ corresponding to species $I$.  
In order for deviations from
the fixed point to decay monotonically in magnitude, 
the magnitudes of the eigenvalues of the matrix of partial derivatives 
in Eq.~(\ref{eq:stab}), ${\bf \Lambda} + {\bf 1}$ where $\bf 1$ is the
$\mathcal{N}$-dimensional unit matrix, must be less than unity. 
The values of the
fecundity $F$ that are used in this work (4 for Model A and 2 for
Model B) were chosen to satisfy this requirement for $\mathcal{N}=1$. 

Since new species are created by mutations, we must also study the
stability of the fixed-point community toward ``invaders." 
Consider a mutant invader $i$. 
Then its multiplication rate, in the limit that 
$n_i \ll n_J$ for all $\mathcal{N}$ 
species $J$ in the resident community, is given by 
\begin{equation}
\frac{n_i(t+1)}{n_i(t)} 
= 
\frac{F}{1 + \exp \left[ -\Delta_i(R,\{n_J^*\}) \right]} 
\;.
\label{eq:inv}
\end{equation}
The Lyapunov exponent, $\ln[ n_i(t+1)/n_i(t) ]$, is the {\it
invasion fitness\/} of the mutant with respect to the resident community 
\cite{DOEB00,METZ92}. 
A characteristic feature of the QSS communities 
observed in both Model A and Model B is that only about 1\%
of the mutants that are separated from the resident community 
by a single mutation (``nearest-neighbor species") have multiplication rates
above unity (and most of those are between 1.0 and 1.1). 
In fact, this is true only to a slightly lesser degree for 
mutants separated by two or three mutations from the resident species 
(``next-nearest neighbors" and ``third-nearest neighbors") \cite{RIKV03}. 
Thus, a string of rather unsuccessful, or at best neutral,
mutations is necessary to bring significant change 
to a QSS community -- a fact that to a large extent accounts for their 
high degree of stability. 

\begin{figure}[t] 
\begin{center}
\vspace*{0.3truecm}
\includegraphics[angle=0,width=.47\textwidth]{ThetafigA.eps} 
\hspace{0.5truecm}
\includegraphics[angle=0,width=.47\textwidth]{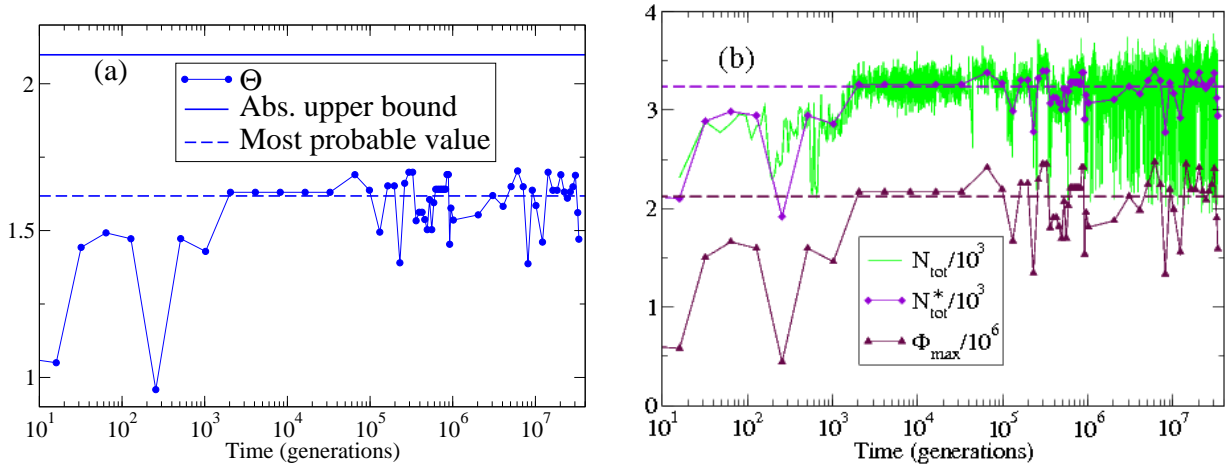} 
\end{center}
\caption[]{
(Color online.) 
Typical time series of the mean-field, fixed-point parameters for Model A
($N_0=2000$, $R=0$, $F=4$, $L=13$, and $\mu = 10^{-3}$), 
emphasizing the transient early-time regime by using a logarithmic time axis. 
{\bf (a)} 
The effective interaction strength $\Theta$. The solid, horizontal line
is the absolute upper bound on $\Theta$ for Model A, $(1+\ln 3) \approx 2.10$, 
while the dashed, horizontal line is the most probable value, 1.618. 
Both are derived in Appendix~\protect\ref{sec:AppX}. 
{\bf (b)} 
The fixed-point population size $N_{\rm tot}^*$ and the corresponding
maximum value of the community 
fitness 
function, $\Phi_{\rm max}$. The
dashed, horizontal lines are the corresponding
values obtained from the most probable
value of $\Theta$. For comparison, the simulated total population size
(sampled on a much finer time scale) is shown light gray (green online)
in the background. 
}
\label{fig:transA}
\end{figure}
\begin{figure}[ht] 
\begin{center}
\vspace*{0.3truecm}
\includegraphics[angle=0,width=.47\textwidth]{ThetaEfig.eps} 
\hspace{0.5truecm}
\includegraphics[angle=0,width=.47\textwidth]{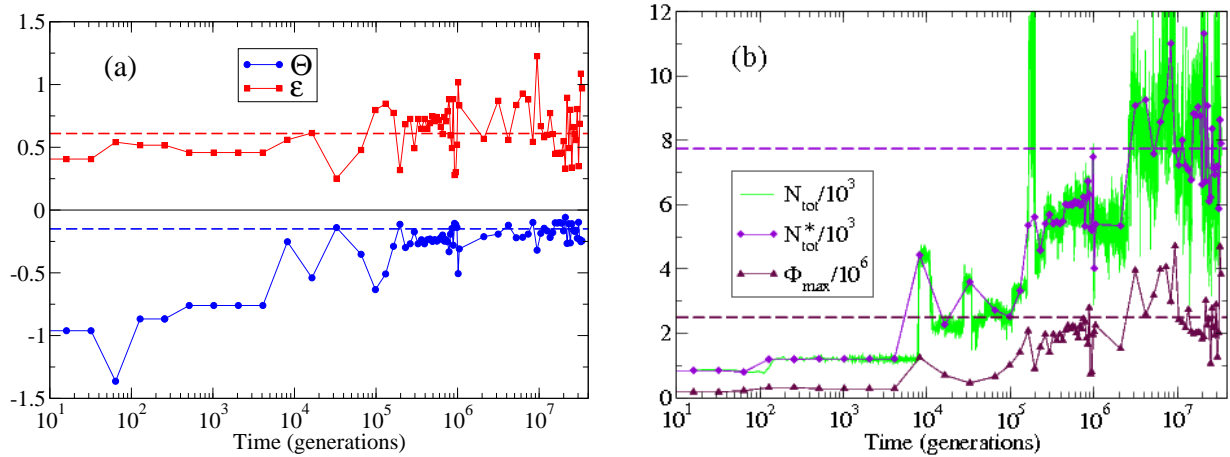} 
\end{center}
\caption[]{
(Color online.) 
Typical time series of the mean-field, fixed-point parameters for Model B
($N_0=\infty$, $R=2000$, $F=2$, $L=13$, and $\mu = 10^{-3}$), 
emphasizing the transient early-time regime by using a logarithmic time axis. 
{\bf (a)} 
The effective interaction strength $\Theta$ and the effective 
coupling to the external resource, $\mathcal{E}$. 
The solid, horizontal line at zero is the absolute upper bound on
$\Theta$ for Model B, 
while the dashed, horizontal lines are the averages over 
late-time communities, also shown as points in Figs.~\protect\ref{fig:ntot}
and~\protect\ref{fig:phi3}. 
{\bf (b)} 
The fixed-point population size $N_{\rm tot}^*$ and the corresponding
maximum value of the community 
fitness 
function, $\Phi_{\rm max}$. 
The dashed, horizontal lines are obtained the same way as in (a). 
For comparison, the simulated total population size
(sampled on a much finer time scale) is shown light gray (green online)
in the background. 
}
\label{fig:transB}
\end{figure}

\section{Comparison of Dynamical Features}
\label{sec:Dyn}

In this section we go beyond the mean-field treatment of the previous 
sections to compare and contrast
some of the dynamical features observed in long 
kinetic Monte Carlo simulations of Model A 
and Model B. For both models we performed multiple 
simulations of $2^{25} = 33\,554\,432$~generations \footnote{The 
simulation length was chosen as a power of two to enable use of the
Fast Fourier Transform algorithm in calculating power spectral densities 
\protect\cite{PRES92}.}.
with a genome of length $L=13$ ($2^{13} = 8192$ potential species)
and a mutation rate per individual of $\mu = 10^{-3}$. The 
fecundity $F$ was set to 4 for Model A and 2 for Model B. 

The model parameters used in our simulations are chosen for
computational feasibility with a view to keeping the system in the
realistic regime where $\mathcal{N}(t) \ll 2^L$ and $\mathcal{N}(t) \ll
N_{\rm tot}(t)$ at all times. At the same time, we want to  
study the dynamics during the late-time, statistically
stationary regime, which should
therefore be reached in a reasonable time. As a result of these
restrictions, $L$, $R$, and $N_0$ are quite small, while $\mu$ is quite
large, compared to natural populations. Results of tests for Model B with
larger $L$ and $R$ and smaller $\mu$ are summarized in Appendix~\ref{sec:AppB}. 
Analogous, but less exhaustive tests for Model A (not reported in detail
here, but see brief summary in Appendix~\ref{sec:AppB}) yield similar results. 
These tests indicate that the products $R \mu$ for Model B and $N_0 \mu$
for Model A are 
proportional to the average number of mutant individuals produced per
generation, $\overline{N_{\rm tot}} \mu$. We also note that this number
is half of
the {\it universal biodiversity number\/} in Hubbell's neutral theory of
biodiversity \cite{HUBB01}. Even though the species in the
models studied here are strongly interacting, this parameter appears to
play a similar role in determining the average diversity
as it does in the neutral theory. Furthermore, the
power laws that are observed and discussed in Sec.~\ref{sec:Dyn2}
are found to be robust, even against parameter changes that change $R \mu$.

\subsection{Evolution of mean-field parameters in the transient regime}
\label{sec:Dyn1}

The simulation runs were started with 100 individuals of
a single species (or of a single producer species for Model B) and allowed to
evolve under the dynamics described in Sec.~\ref{sec:Mod}. At time
points sampled at approximately constant intervals on a logarithmic
scale, the ``stable core" of the community was identified as follows. 
The fixed-point population sizes for the species present in the
community at that time were calculated according to Eq.~(\ref{eq:ssn}),
and species corresponding to negative population sizes were excluded 
(starting from the one with the smallest simulated population size) 
until a stable, feasible community was obtained. This procedure was found to 
essentially correspond to removing species with populations below 100,
which mostly were unsuccessful mutants of the ``core" species. Once the
core community was identified, the corresponding
mean-field parameters $\Theta$,
$\mathcal{E}$ (for Model B only), $N_{\rm tot}^*$, and $\Phi_{\rm max}$
were calculated from Eqs.~(\ref{eq:quad2}) through (\ref{eq:fit}). 

Typical time series of these mean-field parameters are shown in
Figs.~\ref{fig:transA} and~\ref{fig:transB}, in which the early-time
transient regime is emphasized by using a logarithmic time axis. 
For both models a clear tendency is seen for the mean-field parameters to 
increase during the transient regime, for so to fluctuate randomly about
their long-time averages during the the later-time, statistically
stationary regime. Both models thus evolve from an initial
low-population community, toward a sequence of 
QSS communities that optimize their population sizes,
subject to the constraints inherent in the respective models. 

\subsection{Long-time dynamics in the statistically stationary regime}
\label{sec:Dyn2}

Here we concentrate on comparing and contrasting the dynamical properties 
of the two models in the statistically stationary regime that follows
the transient regime discussed above. The emphasis will be on the
temporal fluctuations of the total population sizes and the diversities
of the resulting communities. 
\begin{figure}[t] 
\begin{center}
\vspace*{0.3truecm}
\includegraphics[angle=0,width=.47\textwidth]{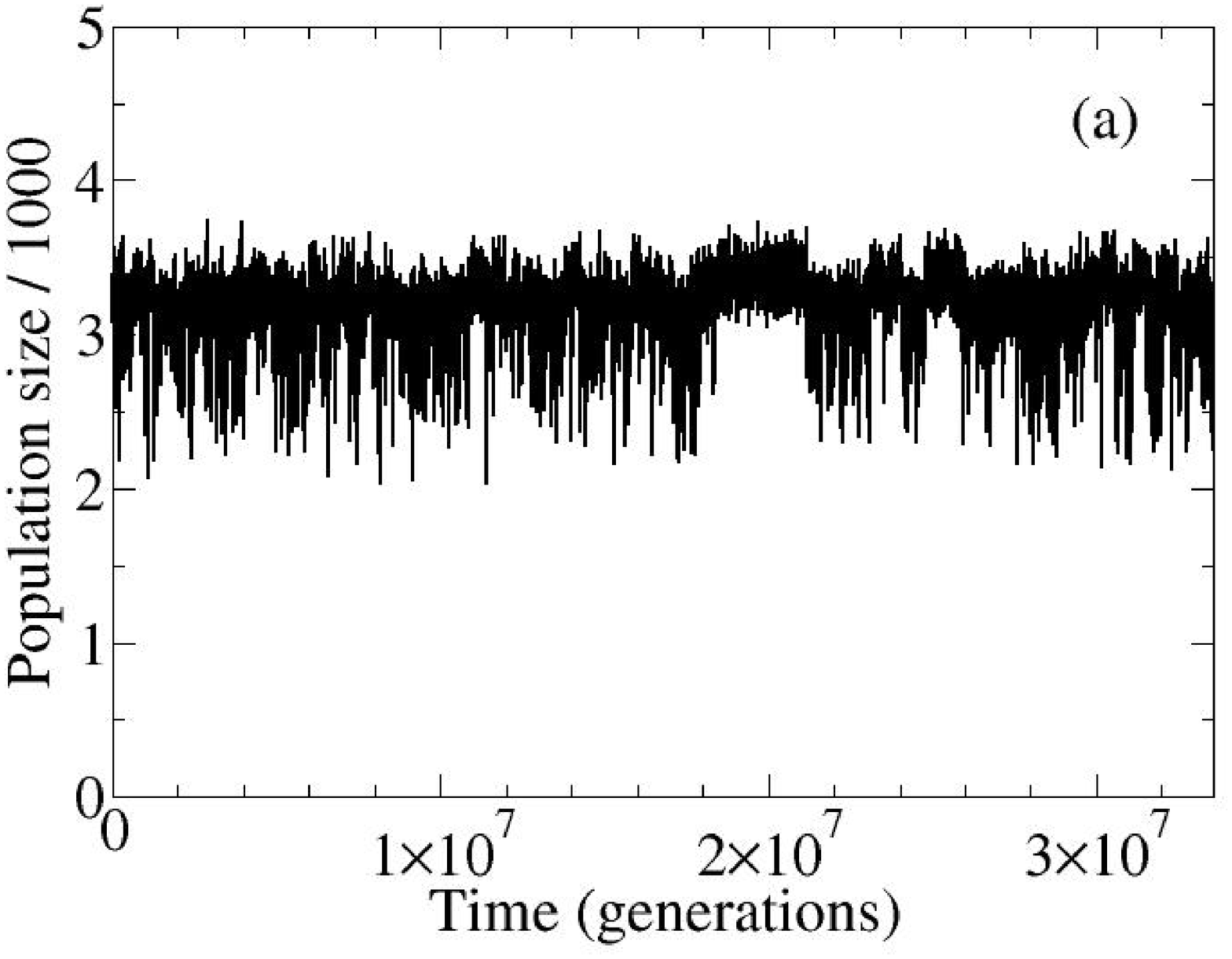} 
\hspace{0.5truecm}
\includegraphics[angle=0,width=.47\textwidth]{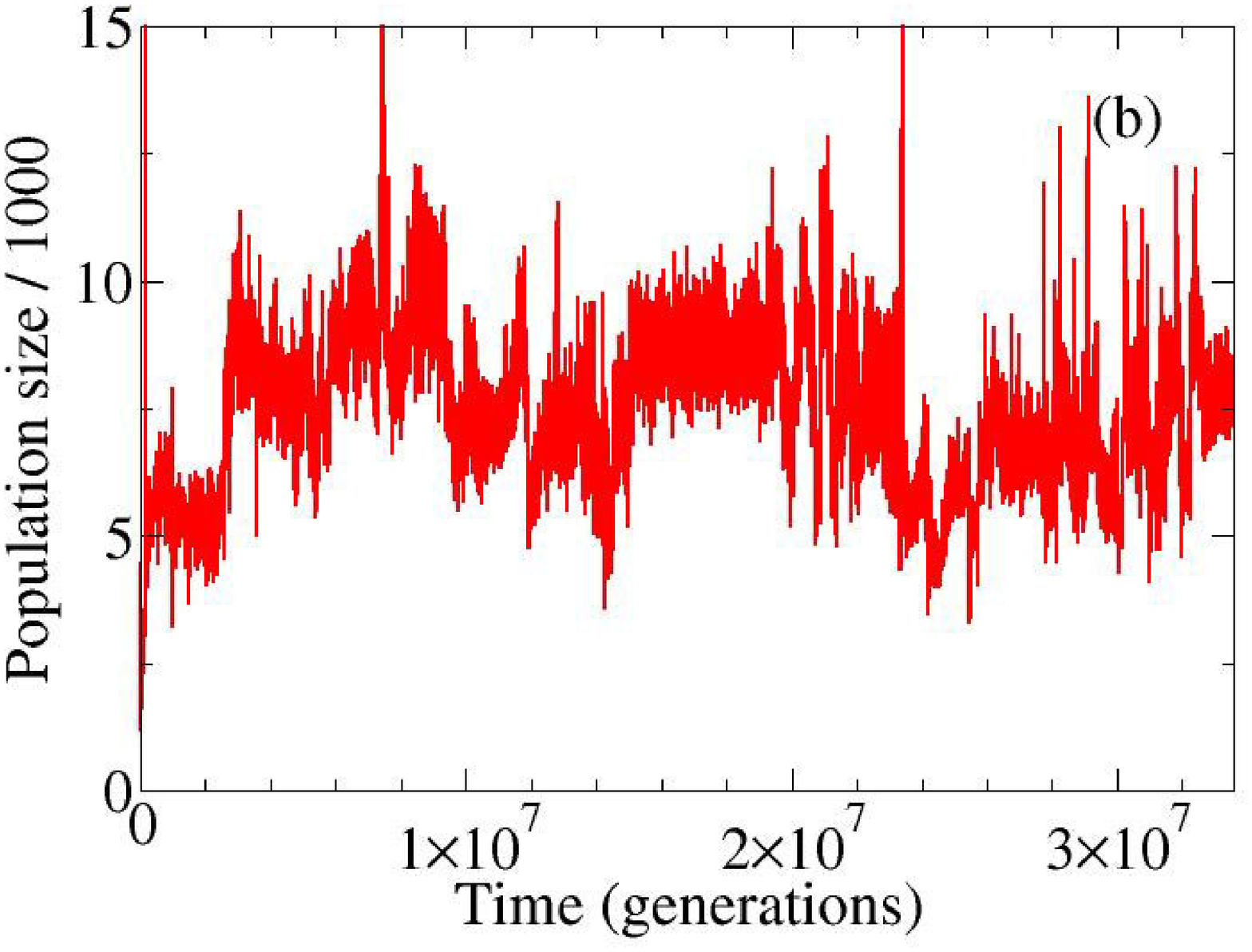} 
\end{center}
\caption[]{
(Color online.) 
Typical time series of the total population sizes, $N_{\rm tot}(t)$, 
for Monte Carlo runs of $2^{25} = 33\,554\,432$ generations each. 
To reduce the size of the figure file, the data are sampled only
once every 2048 generations. While this slightly reduces the
apparent range of the short-time fluctuations, the general shapes of
the time series are preserved. The data are from the same runs shown in
Figs.~\protect\ref{fig:transA} and~\protect\ref{fig:transB}.
{\bf (a)} 
Model A.
{\bf (b)} 
Model B.
}
\label{fig:timeser}
\end{figure}

\begin{figure}[t] 
\begin{center}
\vspace*{0.4truecm}
\includegraphics[angle=0,width=.47\textwidth]{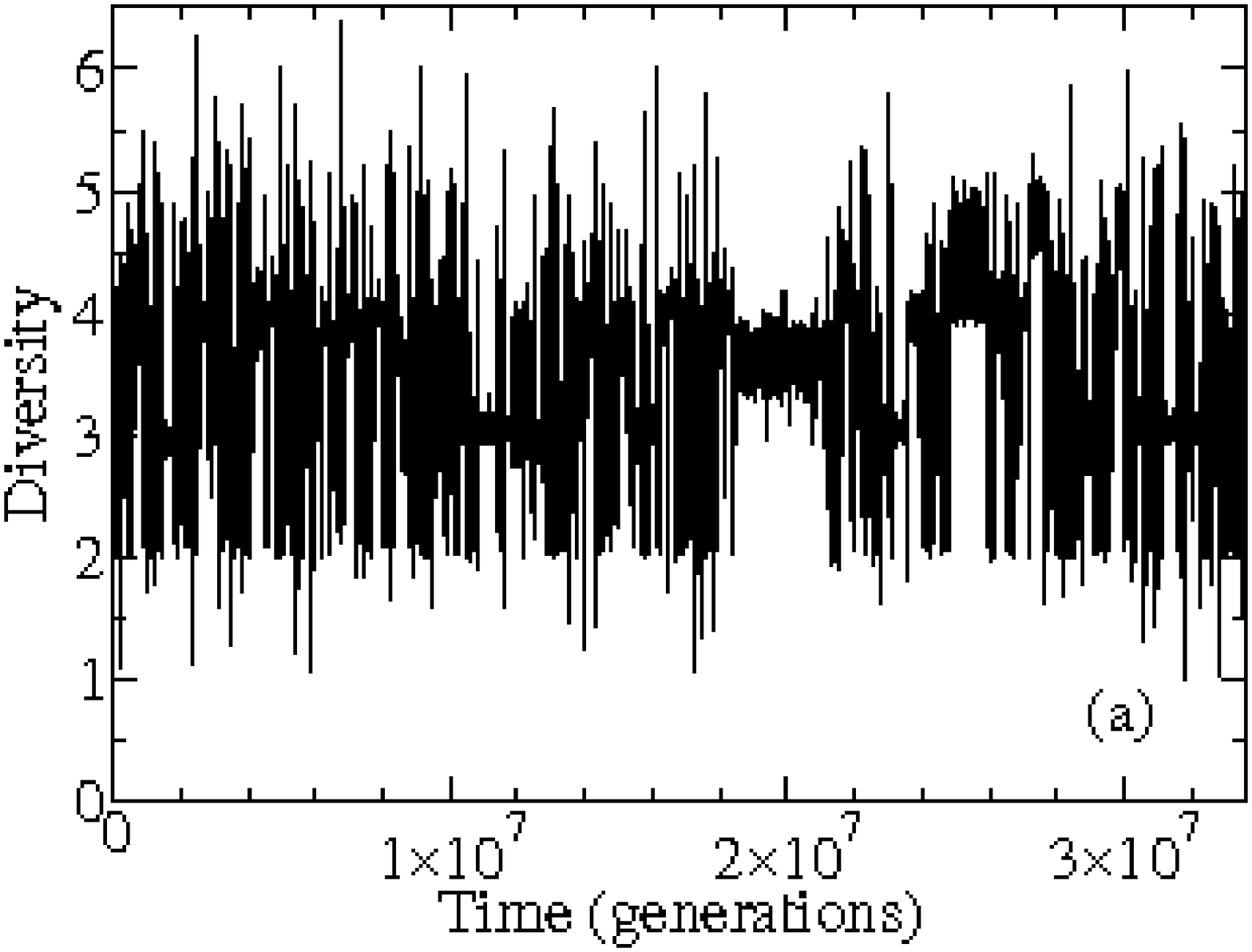} 
\hspace{0.5truecm}
\includegraphics[angle=0,width=.47\textwidth]{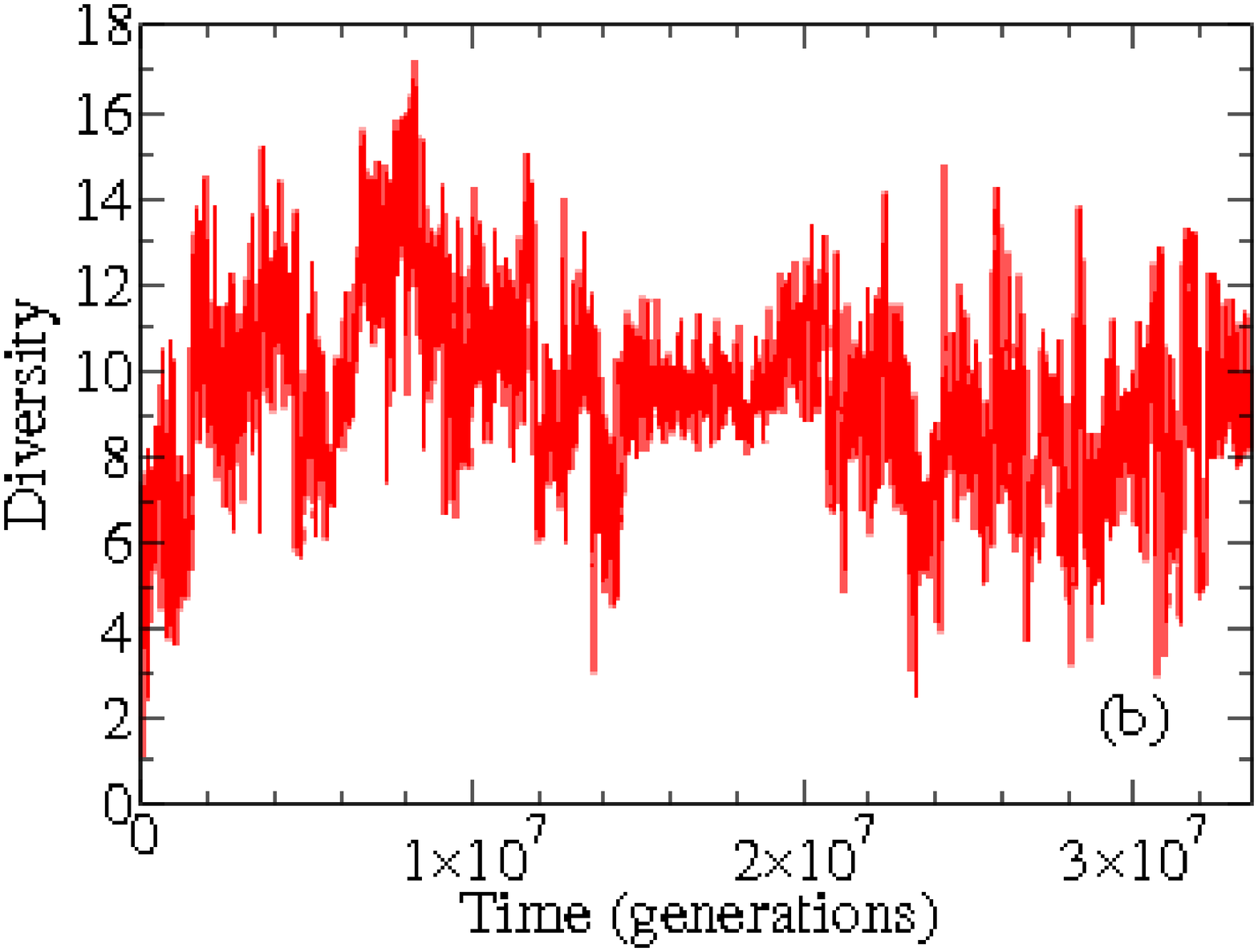} 
\end{center}
\caption[]{
(Color online.) 
Typical time series of the exponential Shannon-Wiener diversity index $D(t)$ 
in the same $2^{25}$-generations Monte Carlo runs shown in 
Fig.~\protect\ref{fig:timeser}. 
The data were downsampled as in Fig.~\protect\ref{fig:timeser}. 
{\bf (a)}
Model A.
{\bf (b)}
Model B.
}
\label{fig:timeserD}
\end{figure}

Typical time series of the total population sizes 
$N_{\rm tot}(t)$ are shown in Fig.~\ref{fig:timeser}. 
They are characterized by QSSs on different time scales,
separated by periods of high evolutionary activity. 
In addition to the total population sizes, 
we also studied the diversities, defined as 
the number of major resident species. In order to obtain an
approximation for the number of major, or ``core" 
species (which can be thought of as the wildtypes in a quasi-species model 
\cite{EIGE71,EIGE88}), we filter out the low-population 
species that are most likely unsuccessful mutants of the 
wildtypes. This is achieved by using the exponential
Shannon-Wiener diversity index 
\cite{KREB89}, 
\begin{equation}
D(t) = e^{S[\{n_I(t)\}]} \;,
\label{eq:div}
\end{equation}
where 
\begin{equation}
S[\{n_I(t)\}] = - \sum_{\{ I | \rho_I(t) > 0 \}} 
\rho_I(t) \ln \rho_I(t) 
\label{eq:ent}
\end{equation}
with $\rho_I(t) = n_I(t)/N_{\rm tot}(t)$ is the
information-theoretical entropy \cite{SHAN48,SHAN49}. 
Typical time series for $D(t)$ in the two models are shown in 
Fig.~\ref{fig:timeserD}. Just like in the time series for the total 
population sizes, the intermittent 
structure consisting of QSSs on different timescales, 
separated by periods of high evolutionary activity, is clearly seen. 

\begin{figure}[t]
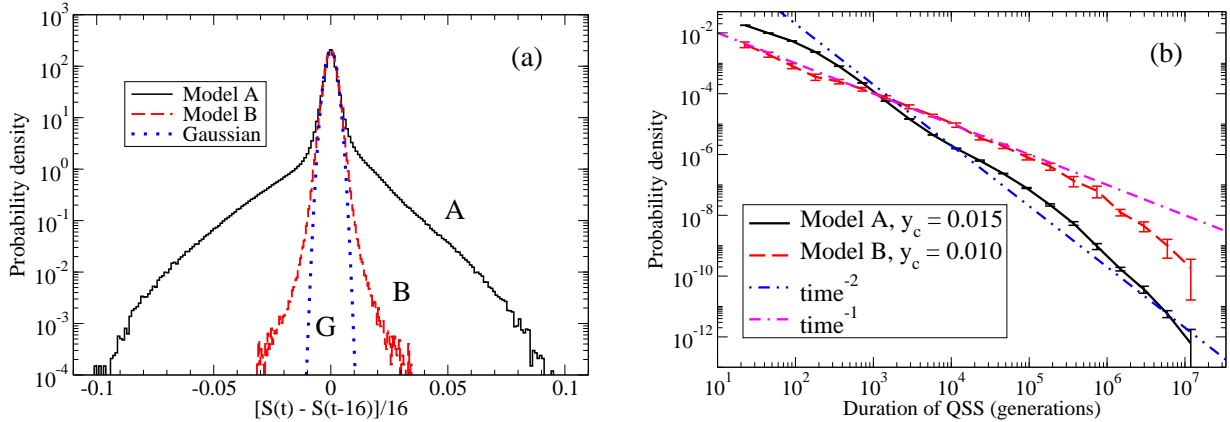
 
\begin{center}
\vspace*{0.3truecm}
\includegraphics[angle=0,width=.47\textwidth]{FigHISTexpS33M.eps} 
\hspace{0.5truecm}
\includegraphics[angle=0,width=.47\textwidth]{FigQuiet_02_DS_talk.eps} 
\end{center}
\caption[]{
(Color online.)
{\bf (a)}
Normalized histograms representing 
the probability density of the logarithmic derivative 
of the diversity, ${\rm d}S(t)/{\rm d}t$. The data were averaged over 16 
generations in each run, and then averaged over 16 independent runs for 
Model A (solid, marked A) and 12 runs for Model B (dashed, marked B). 
The central parts of both histograms are well fitted by a
Gaussian (dotted, marked G).
{\bf (b)}
Log-log plot of normalized 
histograms representing the probability density of the 
durations of QSSs, estimated as
the periods between times when $|{\rm d}S(t)/{\rm d}t|$ exceeds a cutoff $y_c$ 
(${\rm d}S(t)/{\rm d}t$ was averaged over 16 generations as in part (a)). 
Model A (solid) 
and Model B (dashed). Results for Model A were averaged over 16 runs, and 
for Model B over 12 runs. The error bars are standard errors, based on the 
spread between runs. The two dot-dashed straight lines represent time$^{-2}$ 
and time$^{-1}$ power laws, respectively.
}
\label{fig:quiet}
\end{figure}
Statistical information for characteristic time intervals  
that describe the dynamics can be extracted from our data. One such time  
interval is the duration of a QSS. One way to determine 
this is to use a cutoff on the magnitude of the diversity fluctuations, whose 
probability densities for the two models are 
shown in Fig.~\ref{fig:quiet}(a). For both models, the probability densities 
consist of a Gaussian central part representing the fluctuations during the 
QSS periods \cite{ZIA04}, 
flanked by ``wings" that correspond to the large fluctuations during the 
evolutionarily active periods. We choose 
a cutoff $y_c = 0.015$ for Model A and 0.010 for Model B, in both
cases corresponding to the transition region between the Gaussian
central peak and the wings. The 
duration of a single QSS then corresponds to the time interval 
between consecutive times when $|{\rm d}S(t)/{\rm d}t|$ exceeds $y_c$. 
Log-log plots of the resulting probability densities for the durations of 
QSSs in the two models are shown together in Fig.~\ref{fig:quiet}(b). 
While both show approximate power-law behavior over 
five decades or more in time, 
there is an important difference: the exponent 
for Model A is near $-2$, while for Model B it is closer to $-1$. 

\begin{figure}[t]
\begin{center}
\vspace*{0.3truecm}
\includegraphics[angle=0,width=.47\textwidth]{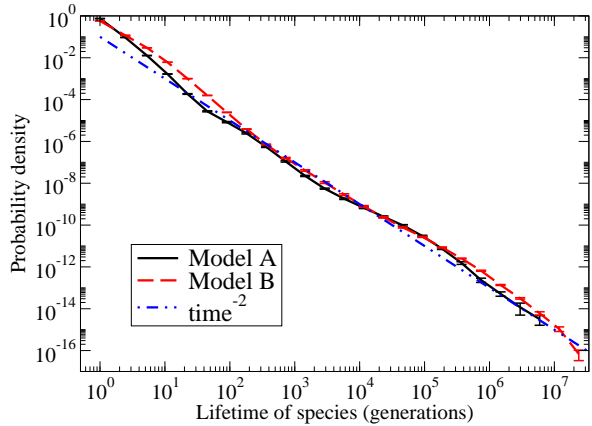} 
\end{center}
\caption[]{
(Color online.)
Log-log plot of normalized 
histograms representing the probability density of the 
lifetime of a particular species. Model A (solid) 
and Model B (dashed). Results for Model A were averaged over eight runs,
and for Model B over 12 runs. 
The error bars were calculated as in Fig.~\protect\ref{fig:quiet}(b). 
The dot-dashed straight line represents a time$^{-2}$ power law.
}
\label{fig:life}
\end{figure}
A different time interval
of interest is the lifetime of a particular species, 
defined as the time between its origination and eventual 
extinction. Log-log plots of histograms of the species lifetimes in the 
two models are shown in Fig.~\ref{fig:life}. In contrast to the case of 
the QSS durations, the species-lifetime distributions are very close for the 
two models, both showing approximate time$^{-2}$ behavior over near 
seven decades in time. 
The observed exponent
is significantly different from $-3/2$, which would correspond to the 
simple hypothesis that the lifetime distributions simply 
correspond to the first-return-time distribution for a random walk of 
$n_I$ \cite{NEWM05,NEWM03}. 
On the other hand, a lifetime distribution with an exponent of $-2$ is
consistent with a stochastic branching process \cite{PIGO05}. 
(Simulations of neutral versions of both models also yield
species lifetime distributions that decay approximately as
time$^{-2}$, but with a sharp cutoff near 10$^4$ generations.)
Lifetime distributions for marine genera that are compatible 
with a power law with an exponent in the range $-1.5$ to $-2$ 
have been obtained from the fossil record \cite{NEWM03,NEWM99B}. 
However, the possible power-law behavior in the fossil record 
is only observed over about one decade in 
time -- between 10 and 100 million years -- and other fitting functions, 
such as exponential decay, are also possible. 
Nevertheless, it is reasonable to conclude that the numerical results 
obtained from complex, interacting evolution models that extend 
over a large range of time scales support interpretations of the fossil 
lifetime evidence in terms of nontrivial power laws. 

\begin{figure}[t] 
\begin{center}
\vspace*{0.3truecm}
\includegraphics[angle=0,width=.47\textwidth]{species_33M_S.eps} 
\hspace{0.5truecm}
\includegraphics[angle=0,width=.47\textwidth]{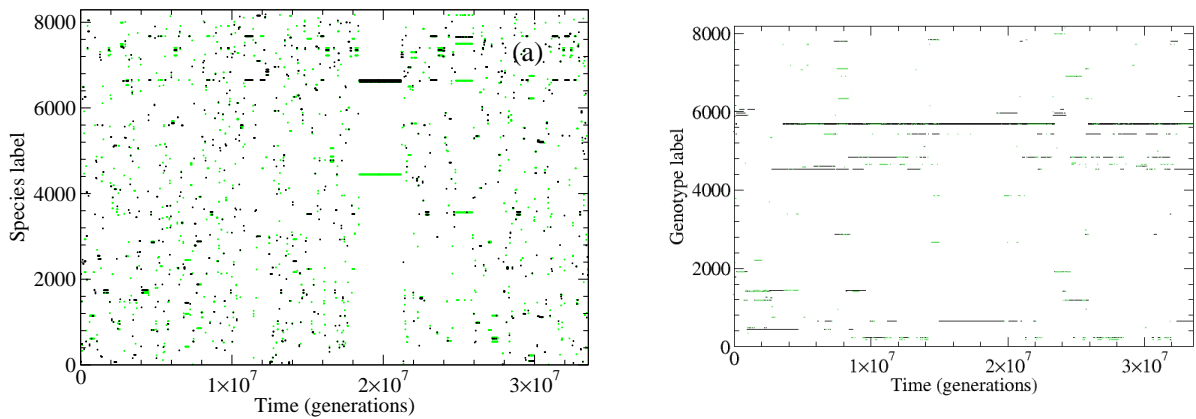} 
\end{center}
\caption[]{
(Color online.) 
Plots showing the labels $I$ of the most highly populated species vs time. 
The data are for the same $2^{25}$-generation Monte Carlo runs shown in
Figs.~\protect\ref{fig:transA}--\protect\ref{fig:timeserD}. 
Gray (green online): $n_I \in [101,1000]$. 
Black: $n_I \ge 1001$. 
{\bf (a)} Model A. 
{\bf (b)} Model B. For clarity, only producer species are shown 
for this model. The consumer species give a similar picture. 
See further discussion in the text. 
}
\label{fig:specs}
\end{figure}
\begin{figure}[t] 
\begin{center}
\vspace{0.5truecm}
\includegraphics[angle=0,width=.47\textwidth]{species_33M_S_magn.eps} 
\hspace{0.5truecm}
\includegraphics[angle=0,width=.47\textwidth]{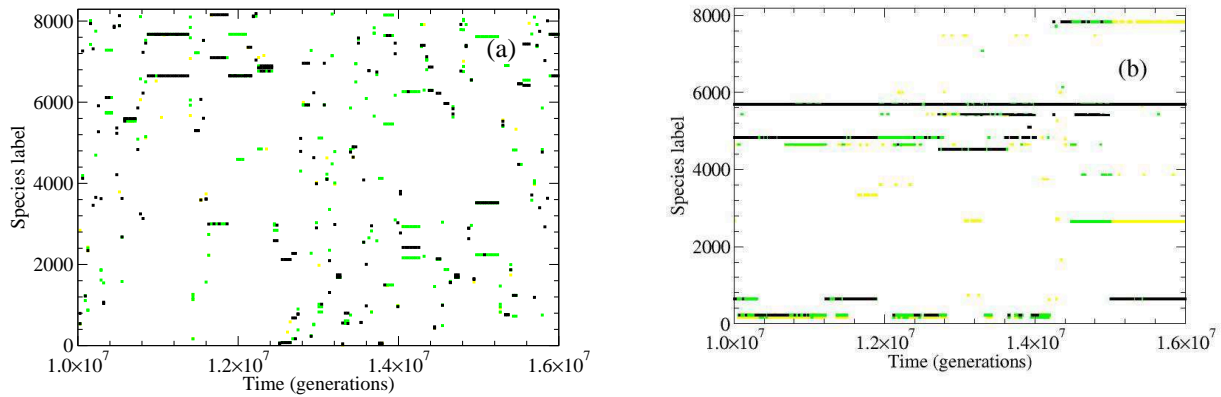} 
\end{center}
\caption[]{
(Color online.) 
Magnified version of Fig.~\protect\ref{fig:specs}, showing times
between $1.0\times 10^7$ and $1.6\times 10^7$ generations. 
In addition to the species shown in Fig.~\protect\ref{fig:specs},
the ones with $n_I \in [11,100]$ are also shown (light gray,
yellow online). 
{\bf (a)} Model A. 
{\bf (b)} Model B. 
See further discussion in the text. 
}
\label{fig:specs2}
\end{figure}
\begin{figure}[t] 
\begin{center}
\vspace*{0.3truecm}
\includegraphics[angle=0,width=.47\textwidth]{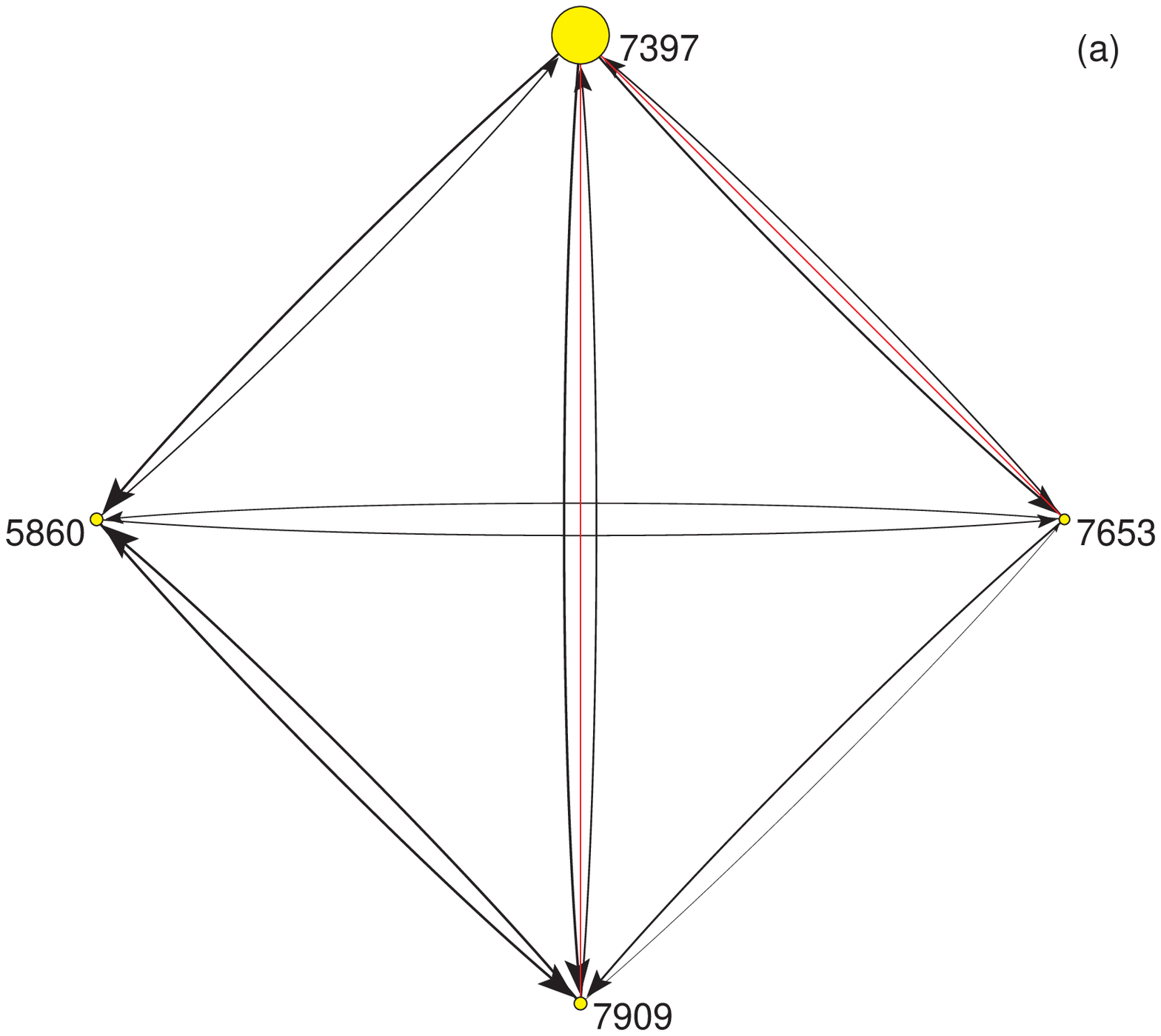} 
\hspace{0.5truecm}
\includegraphics[angle=0,width=.47\textwidth]{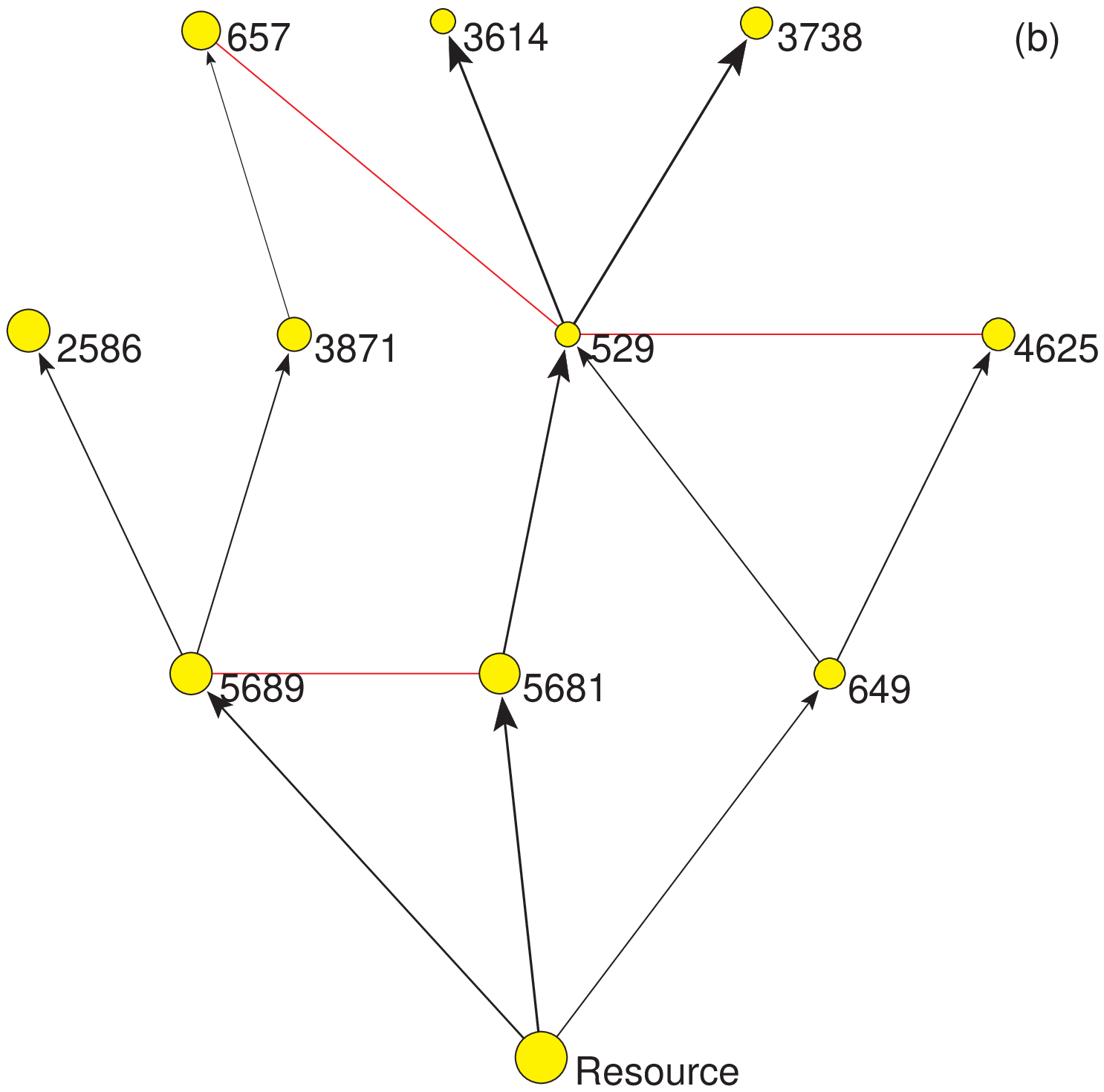} 
\end{center}
\caption[]{
(Color online.) 
Typical QSS core communities observed in simulations of the two models.
The numbers are the species labels $I$, and the sizes of the circles 
represent the individual species' fixed-point populations, 
obtained from Eq.~(\protect\ref{eq:ssn}). 
A black arrow pointing toward species $I$ indicates a positive
$M_{IJ}$, with magnitude indicated by the line thickness and size
of the arrow head. The thin gray lines without arrowheads 
(red online) connect nearest neighbors in genotype space. 
(a) 
Model A. This particular QSS community is the one included in
Table 2 of \protect\cite{ZIA04} and line 10 of Table I of 
\protect\cite{RIKV03}. 
(b)
Model B. This is the food web representing the
QSS community existing between 15 and 18 million
generations in the simulation shown in
Figs.~\protect\ref{fig:transB}, \protect\ref{fig:timeser}(b), 
\protect\ref{fig:timeserD}(b),  
\protect\ref{fig:specs}(b), and~\protect\ref{fig:specs2}(b). 
It has three trophic levels above the resource node. 
}
\label{fig:Webs}
\end{figure}

The difference in the power laws for the QSS durations and the species 
lifetimes is a puzzling result. A likely explanation can be gleaned 
from the data shown in Figs.~\ref{fig:specs} and~\ref{fig:specs2}. 
These show the species labels 
of highly populated species as functions of time for both models. From 
Figs.~\ref{fig:specs}(a) and~\ref{fig:specs2}(a)
it is seen that all or most of the horizontal lines representing
populated species at a given time for Model A start
and stop almost simultaneously, indicating that species originations
and extinctions in this model are highly synchronized.  
In other words: whole communities in Model A tend to go extinct and be
replaced with an entirely new community within a short time. 
In contrast, in Figs.~\ref{fig:specs}(b) and~\ref{fig:specs2}(b) 
the different horizontal species 
lines for Model B stop and start at different times. This indicates that 
communities in this model are much more robust, and extinction 
events seldom wipe out more than a part of the total community. Thus, QSSs 
would be expected to be more long-lived (but also less clearly defined) 
for Model B, than for Model A. This theory is supported by the
strikingly different QSS community structures produced by the two
models, shown in Fig.~\ref{fig:Webs}. The typical QSS community
shown for Model A is a small cluster of mutualistically interacting
species, while the typical community shown for Model B has the
character of a simple food web. Extinctions in the latter are
likely to be confined to a single branch of the web. 
A detailed discussion of the community structures generated by Model B,
including comparisons with data from real food webs, is found in \cite{RIKV06B}. 

\begin{figure}[t]
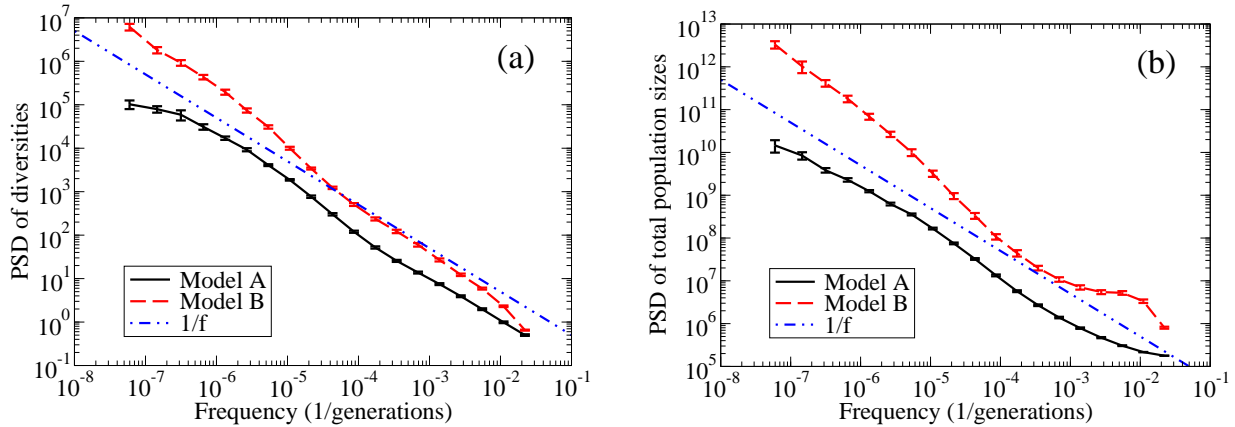
 
\begin{center}
\vspace*{0.3truecm}
\includegraphics[angle=0,width=.47\textwidth]{AvPSD_fig_talk.eps} 
\hspace{0.5truecm}
\includegraphics[angle=0,width=.47\textwidth]{AvPSD_pop_fig_talk.eps} 
\end{center}
\caption[]{
(Color online.) 
Log-log plots of 
PSDs of {\bf (a)} the diversities $D(t)$ and {\bf (b)} 
the total population sizes $N_{\rm tot}(t)$ for Model A and Model B. 
The PSDs are averaged over each octave in frequency, and then 
over 16 runs for Model A and 12 runs for Model B. 
The error bars were calculated as in Fig.~\protect\ref{fig:quiet}(b). 
The dashed straight lines represent $1/f$ power-law behavior.
}
\label{fig:PSD}
\end{figure}

The arbitrariness inherent in the cutoff that must be used to extract 
QSS duration distributions from fluctuations in the diversity or other time 
series can to some extent be eliminated by mapping distributions obtained 
with different cutoffs onto a common scaling function 
\cite{PACZ96,RIKV05A}. However, an analysis method that completely 
avoids cutoffs is that of calculating power spectral densities 
(the square of the temporal Fourier transform). Power spectra 
(PSDs) are therefore shown in 
Fig.~\ref{fig:PSD} for both models. The PSDs for the diversity are 
shown in Fig.~\ref{fig:PSD}(a), and for the total population size in 
Fig.~\ref{fig:PSD}(b). 
Although there are clear deviations, the overall behaviors for both quantities 
and for both models are compatible with a $1/f$ power law over many
decades in frequency. In the 
high-frequency regime the population-size PSDs have a significant background of 
noise, presumably caused by the rapid population fluctuations due to the 
birth and death of individual organisms. 
For very low frequencies there is little reason to believe that there should 
be large differences between the behaviors of the two quantities for the same 
model. We therefore think it is reasonable to consider the difference between 
the slopes of 
the diversity and population-size PSDs as an indication of the true 
uncertainty in the PSDs at the lowest frequencies. Better estimates in this 
regime would require several orders of magnitude longer simulations. 

\section{Discussion and Conclusions}
\label{sec:Conc}

In this paper we have compared and contrasted the long-time evolutionary
dynamics of two individual-based models of biological coevolution,
using both analytic linear stability calculations and large-scale
kinetic Monte Carlo simulations. 
These models involve universal competition and ignore 
important effects such as adaptive foraging. They are therefore not 
highly realistic. However, the fact that the results of numerical 
simulations can be compared with exact 
analytical results make these simplified 
models ideal as benchmarks for simulations of more realistic 
coevolution models in the future \cite{RIKV06C}. 

A central result of the analytic study is that, in the
absence of mutations, the total 
population size of a fixed-point community, $N_{\rm tot}^*$, 
is described by a community 
fitness 
function that is cubic in $N_{\rm tot}$. 
In addition to the present application, such a model
is also applicable to nonequilibrium phase transitions in 
such diverse systems as epidemics, 
lasers, and autocatalytic chemical reactions.  However, these evolution 
models differ from those kinds of systems by the 
important effect that, in the {\it presence\/} of mutations, 
the model parameters $\Theta$ and $\mathcal{E}$ are no longer 
externally imposed constraints, but rather {\it evolve\/} 
as far in the direction of positive $\Theta$ and large $\mathcal{E}$ 
as allowed by the internal constraints of the particular model. 
As a result, Model A, in which the elements of the 
interspecies interaction matrix $\bf M$ are randomly distributed on 
an interval that is symmetric about zero, 
evolves to produce communities that 
are heavily biased toward mutualism. The effective interaction variable 
$\Theta$ adjusts to a positive value.
In contrast, the predator-prey Model B, in which the 
interspecies interactions 
are antisymmetric, is constrained to nonpositive values of 
$\Theta$, for which a nonzero population size can 
only be sustained through the external resource $R$. 
These results are illustrated in Fig.~\ref{fig:ntot}(a). 

In a recent study of a similar, but different model for coevolution 
\cite{TOKI03}, the emergence of strongly mutualistic 
communities from initially unbiased conditions was also observed. 
In that model, mutants 
are very similar to their parents, except for their interactions with a 
few other species (``local mutations"), 
and the authors suggest that the evolution of mutualism 
is related to this feature of their model. 
However, the mutations in our models would be ``global" in the language of 
\cite{TOKI03}, which leads us to the conclusion that the emergence of 
mutualism is common in models where direct 
mutualistic interactions are allowed. 
Rather remarkably, we have found little difference in the dynamics between 
the version of Model A studied in this paper and a version with 
strongly correlated interactions, and thus more ``local" mutations
\cite{SEVI06}. 
It remains an important problem to reconcile this tendency for evolution 
of mutualism with the obvious requirement that biomass cannot be created 
without energy input. 
While predator-prey interactions are easy to reconcile 
with energy conservation, direct mutualistic interactions 
are more difficult to interpret in an energy framework \cite{BRON94}
(although effective mutualism is  
common in nature \cite{BASC06,BRON94,KAWA93,KREB01}). 
We believe this emphasizes the importance of distinguishing between
{\it direct\/} mutualistic interactions, as in Model A, and more realistic 
{\it effective\/} mutualisms, which merely mean that a pair of elements in
the community matrix, $\Lambda_{IJ}$ and
$\Lambda_{JI}$, are both positive.  
The unrealistic emergence of direct mutualism in Model A is shared by
the Tangled-nature model \cite{CHRI02,COLL03,HALL02}, 
of which it is a simplified version. 

Beyond the mean-field studies and the simulation results for average 
population sizes, 
we have also studied the temporal fluctuations of both the 
diversity and total population size for Models A and B. We find that 
the probability distributions of the lifetimes of individual species in 
both models are very similar, 
showing power-law decay with an exponent near 
$-2$ over near seven decades in time, as seen in Fig.~\ref{fig:life}. 
This exponent value is consistent 
with some interpretations of the available 
data for the lifetimes of marine genera 
in the fossil record \cite{NEWM03,NEWM99B}, 
but other interpretations of the fossil evidence are also possible. 
Similarly, power spectra for the diversity, as well as for the total
population size, 
show reasonable (although not perfect) $1/f$ behavior over many decades  
in frequency, as seen in Fig.~\ref{fig:PSD}. 
It is therefore very interesting that the probability distributions of the 
durations of individual QSS periods in the two models also both show 
reasonable power-law decay, but with {\it different\/} exponents: 
near $-2$ for Model A and close to $-1$ for Model B, as seen in 
Fig.~\ref{fig:quiet}(b). 
This result, which we found quite surprising at first, makes sense in 
light of the observation that the extinctions of major species are 
highly synchronized in Model A, while they are much less so in Model B. 
While communities in Model A tend to collapse completely when 
an aggressive mutant arrives and/or a major species goes extinct, 
communities in Model B are much more resilient and extinctions most often 
only extend to one or a few branches 
of the resident food web. This effect is illustrated in 
Figs.~\ref{fig:specs}, \ref{fig:specs2}, and \ref{fig:Webs}. 
The long-time correlations that give rise to these extended power
laws are clearly due to the interspecies interactions. Indeed, our
exploratory simulations of neutral versions of both models (not shown here) 
exhibit no correlations beyond about 10$^4$ generations. 

Our observation of the high resilience of Model B against complete  
extinction of communities is consistent with observations of 
extinction avalanches of limited size in the Web-world model by
Drossel et al.\  
\cite{DROS01B}, who argue that their model is 
therefore {\it not\/} self-organized 
critical. Together with our observation of the self-optimization of 
the evolution models studied here to points {\it away from\/} 
the transcritical bifurcation point, these observations may support a 
conclusion that models of coevolution that take reasonable account of the 
dynamics at the ecological level (even if they are extremely simplified) 
are not in general self-organized critical. Such a conclusion, which
appears reasonable, would be in 
disagreement with a number of recent theories of extinction 
\cite{BAK93,NEWM03,PACZ96}. 

The discussion in the preceding paragraphs brings out some of the
differences and similarities between the models studied here, and some
of the other physics-inspired evolution models that have been introduced
in recent years. Like the Tangled-nature model \cite{CHRI02,COLL03,HALL02}, 
our models are individual-based, and Model A also shares with that model
the unrealistic emergence of mutualistic communities. The power-law
behaviors are also reminiscent of species-based (rather than
individual-based) extremal-dynamics evolution
models like the Bak-Sneppen model and its generalizations 
\cite{BAK93,NEWM03,PACZ96}, and in \cite{RIKV05A} we speculated
that Model B might correspond to a zero-dimensional extremal-dynamics
model \cite{DORO00}. However, more accurate exponent estimates for Model
B \cite{RIKV06B}, as well as the above comparison of Model B with the 
Web-world model \cite{DROS01B}, make this conjecture less likely to hold. 

In conclusion, the results presented here indicate that further work on 
models of macroevolution that are based on events on ecological time scales, 
with comparisons of the results with 
data from the fossil record, as well as from laboratory experiments and 
extant food webs, is 
highly desirable. In a separate paper we consider in detail the 
structure and dynamics of the food webs that develop within Model B, 
and we compare these with data for real food webs \cite{RIKV06B}.

\begin{acknowledgements}
The author thanks R.~K.~P.\ Zia, B.~Schmittmann, P.~Beerli, 
G.~J.~P.\ Naylor, and V.~Sevim 
for useful discussions and comments on the manuscript, 
and V.~Sevim for the data on Model A included in Fig.~\ref{fig:life}. 
He also gratefully acknowledges hospitality at Kyoto University 
by H.~Tomita in the 
Department of Fundamental Sciences, Faculty of Integrated Human Studies,
and H.~Fujisaka in the 
Department of Applied Analysis and Complex Dynamical Systems, 
Graduate School of Informatics. 

This work was supported in part by U.S.\ National Science Foundation Grants 
No.\ DMR-0240078 and DMR-0444051, and by Florida State University
through the School of Computational Science, the Center for Materials
Research and Technology, the National High Magnetic Field Laboratory,
and a COFRS summer-salary grant.
\end{acknowledgements}

\appendix

\section{Derivation of the Community Fitness Function}
\label{sec:AppA}

In this Appendix we provide a conventional derivation of the cubic 
form of the community 
fitness 
function $\Phi(N_{\rm tot})$ in a simple 
mean-field approximation. The derivation provides an explanation for the
prefactor $(1 - 1/F)$ in Eq.~(\ref{eq:fit}), as well as intuitively
clear approximations for the coefficients $\Theta$ and $\mathcal{E}$. It
also provides justification that the equation to be integrated
to obtain $\Phi(N_{\rm tot})$ is indeed Eq.~(\ref{eq:quad}), rather than
this equation multiplied or divided by some power of $N_{\rm tot}$. 
The derivation is based on the time-dependent Ginzburg-Landau equation
for a system with nonconserved order parameter 
\cite{GOLD92,HOHE77}, which for our current systems takes the form 
\begin{equation}
\frac{\partial N_{\rm tot}}{\partial t} 
=
\frac{\partial \Phi}{\partial N_{\rm tot}} 
\;.
\label{eq:tdgl}
\end{equation}

Identifying $\partial n_I(t) / \partial t$ with 
$n_I(t+1) - n_I(t)$, we obtain from Eqs.~(\ref{eq:PI}--\ref{eq:MF}) 
in the absence of mutations: 
\begin{equation}
\frac{\partial n_I(t)}{\partial t}
=
n_I(t) \left\{
\frac{F}
{1 + \exp \left[ b_I - \frac{\eta_I R}{N_{\rm tot}(t)} 
- \sum_J M_{IJ} \frac{n_J(t)}{N_{\rm tot}(t)} + \frac{N_{\rm tot}(t)}{N_0} 
\right]}
-1 \right\}
.
\label{eq:dndt}
\end{equation}
Expanding this nonlinear equation of motion around its fixed point, we get
\begin{equation}
\frac{\partial n_I(t)}{\partial t}
\approx
\left( 1 - \frac{1}{F} \right) 
n_I(t) 
\left[ - \tilde{b}_I + \frac{\eta_I R}{N_{\rm tot}(t)} 
+ \sum_J M_{IJ} \frac{n_J(t)}{N_{\rm tot}(t)} - \frac{N_{\rm tot}(t)}{N_0} 
\right]
.
\label{eq:dndt2}
\end{equation}
To obtain the simplest mean-field approximation for ${\partial N_{\rm
tot}}/{\partial t}$ (exact for $\mathcal{N}=1$), we set 
$n_I \approx N_{\rm tot}/\mathcal{N}$, 
$\tilde{b}_I \approx \mathcal{N}^{-1}\sum_I' \tilde{b}_I 
\equiv \langle \tilde{b} \rangle$, 
$\eta_I \approx \mathcal{N}^{-1}\sum_I' \eta_I \equiv \langle \eta \rangle$, 
and 
$M_{IJ} \approx \mathcal{N}^{-2} \sum_{IJ}' M_{IJ} \equiv \langle M
\rangle $, where the primes on the sums indicate that they are
restricted to the $\mathcal{N}$ species with $n_I > 0$. 
This yields 
\begin{equation}
\frac{\partial N_{\rm tot}}{\partial t} 
\approx 
\left( 1 - \frac{1}{F} \right) 
\left[
R \langle \eta \rangle  
+ \left( \langle M \rangle - \langle \tilde{b} \rangle \right) N_{\rm tot}
- \frac{(N_{\rm tot})^2}{N_0}
\right]
.
\label{eq:dNdt}
\end{equation}
Integrating the right-hand side as prescribed by Eq.~(\ref{eq:tdgl}), 
we find 
\begin{equation}
\Phi(N_{\rm tot}) 
\approx 
\left( 1 - \frac{1}{F} \right) 
\left(
R  \langle \eta \rangle N_{\rm tot} 
+ \frac{\langle M \rangle - \langle \tilde{b} \rangle}{2} N_{\rm tot}^2
- \frac{1}{3 N_0}N_{\rm tot}^3
\right)
.
\label{eq:PhiMFA}
\end{equation}
This result has the same cubic form as Eq.~(\ref{eq:fit}), with the
approximate coefficients 
$\langle \eta \rangle \approx \mathcal{E}$ and 
$(\langle M \rangle - \langle \tilde{b} \rangle ) \approx \Theta$. 
The exact fixed-point solution for $N_{\rm tot}$ requires the use of 
$\mathcal{E}$ and $\Theta$, but the approximate forms 
obtained here may provide 
a more intuitive understanding of the significance of these coefficients.

\section{Derivation of Optimum Point for Model A}
\label{sec:AppX}

We can obtain a theoretical estimate for the point 
($\overline{\Theta},\overline{N_{\rm tot}})$ in Model A. 
We assume for simplicity that all the off-diagonal $M_{IJ}$
have the same value, $a$. Then, using the definition of $\Theta$, 
and remembering that $F=4$ for Model A, so that 
$\tilde{b}_I = -\ln 3$ for all $I$, one can show
that $\Theta = (1 - 1/\mathcal{N})a + \ln 3$, where
$\mathcal{N}$ is the total number of species in the community. 
This yields the absolute maximum value for $\Theta$
equal to $(1+\ln 3) \approx 2.10$ 
for $\mathcal{N} = \infty$ (shown by vertical
full lines in Fig.~\ref{fig:ntot}). However, it has been 
shown \cite{RIKV03} that the most probable number of species in a
community within this model is finite and given by
\begin{equation}
\mathcal{N}^\dag \approx L \ln2 /\ln (1/q) \;,
\label{eq:Ndag}
\end{equation}
where $q$ is the probability of finding a pair of interactions, 
$M_{IJ}$ and $M_{JI}$, conducive to this community. 
With the $M_{IJ}$ independently distributed on $[-1,+1]$, 
the probability of drawing a
pair that are both larger than some given value $m \in [-1,+1]$, 
is $q = [(1-m)/2]^2$.  In our approximate formula for 
$\Theta$, we now replace $\mathcal{N}$ 
by $\mathcal{N}^\dag$ with this value of $q$, and $a$ by 
the average of a variable uniformly distributed over $[m,1]$, 
which is $(1+m)/2$. This yields 
\begin{equation}
\Theta 
\approx
\left[ 
1 - \frac{2 \ln \left( \frac{2}{1-m} \right)}{13 \ln 2}
\right] 
\frac{1+m}{2} + \ln 3 
\label{eq:1x1}
\end{equation}
for $L=13$.
This is a concave function of $m$ with a single maximum at $m_{\rm
max}$, which can be found numerically. The result is
$m_{\rm max} \approx 0.512$, which yields 
$\Theta_{\rm max} \approx 1.618$, 
$N_{\rm tot} \approx 3236$, and $\overline{M_{IJ}} = (1+m_{\rm max})/2
\approx 0.756$. These values are in excellent agreement with those
obtained from the Monte Carlo simulations, and they are
shown as gray dots (turquoise online) in Fig.~\ref{fig:ntot}.

\section{Trials with Other Parameter Values}
\label{sec:AppB}

In addition to the parameter set used in the main simulations
presented above, we also performed trial simulations for Model B
with $L=20$ (1\,048\,576 potential species), $\mu = 2.5 \times
10^{-4}$, and $R = 8000$. As each simulation run with $R=2000$ took
about two weeks of CPU time, and each run with $R=8000$ took at
least four weeks, relatively few trial runs were performed.
No qualitative differences were observed
in the quantities reported on the basis of the main simulation series. 

To obtain estimates of the effects of $L$, $\mu$, and $R$ on the
the mean stationary levels of the total
population size, $\overline{N_{\rm tot}}$, and diversity, $\overline{D}$, 
and the mean time to reach stationarity, $\overline{\tau}$, 
we fitted these variables by the
exponential function $a[1 - \exp(-t/\tau)]$.
Numerical results for $\overline{\tau}$, 
$\overline{D}$, and $\overline{N_{\rm tot}}$, based on 
twelve runs for 
$L=13$, $R=2000$, and $\mu = 10^{-3}$ 
(the original runs used in the main part of this paper);
twelve runs for 
$L=20$, $R=2000$, and $\mu = 10^{-3}$; 
nine runs for 
$L=20$, $R=2000$, and $\mu = 2.5 \times 10^{-4}$; and five runs for 
$L=20$, $R=8000$, and $\mu = 2.5 \times 10^{-4}$ are compiled in
Table~\ref{table:I}. 

\subsection{Effects of the genome size}

With $L=13$, the whole pool of potential species is visited in a
typical $2^{25}$-generation simulation, while only about 10\% 
are typically visited for $L=20$ with $\mu=10^{-3}$ and $R=2000$. However, the 
``revivals" of extinct species, seen frequently with $L=13$
\cite{RIKV03} and much less
so with $L=20$, has no effect on the observed power laws. 
Twelve full runs were performed for this parameter set. 
All the subsequent trial simulations were performed with $L=20$. 

As expected from Eq.~(\ref{eq:Ndag}), $\overline{D}$ appears to be
proportional to $L$. The significant increase in $\overline{N_{\rm tot}}$ 
when $L$
is increased from 13 to 20 (not seen in trial runs with $L=20$ for Model A) 
is probably due to the
increased probability of finding species with smaller $b_I$ and thus
closer to the population divergence for Model B at $\Theta = 0$ 
(see Fig.~\ref{fig:ntot}(a)). 

\begin{table}[t]
\caption[]{Numerical results for the main simulations of Model B with 
$L=13$, compared with trial runs with $L = 20$ and varying $R$ and $\mu$.
The unit for $R$ and $\overline{N_{\rm tot}}$ is individuals, for
$\mu$ it is mutations per individual offspring per generation, for 
$\overline{\tau}$ it is $10^6$ generations, and for $\overline{D}$ it is
species. Thus, $R \mu$ is proportional to the total number of
mutated offspring per generation. The uncertainties are standard
errors, based on the spread between individual runs.  
$\sigma_\tau$ is the standard deviation of $\tau$ over the
independent runs. A ratio $\overline{\tau}/\sigma_\tau \approx 1$
indicates that $\tau$ may be approximately exponentially distributed, 
while the even lower ratio for $R\mu = 0.5$ indicates 
very large variations from run to run. 
}
\begin{center}
\begin{tabular}{|c|c|c|c|c|c|c|c|c|} 
\hline
$L$ & $R$ & $\mu$ & $R \mu$ & Runs & $\overline{\tau}$ & 
$\overline{\tau}/\sigma_\tau$ & $\overline{D}$ 
& $\overline{N_{\rm tot}}$  
\\
\hline\hline
13 & 2000 & $10^{-3}$ & 2.0 & 12 & 0.35$\pm$0.07 & 1.08 & 10.9$\pm$0.3 
& 7\,700$\pm$300 
\\
\hline
20 & 2000 & $10^{-3}$ & 2.0 & 12 & 0.8$\pm$0.2 & 0.93 & 14.7$\pm$0.3 
& 13\,700$\pm$400 
\\
\hline
20 & 8000 & $2.5 \times 10^{-4}$ & 2.0 & 5 & 1.2$\pm$0.3 & 0.90 & 20.4$\pm$0.6 
& 58\,000$\pm$7\,000 
\\
\hline
20 & 2000 & $2.5 \times 10^{-4}$ & 0.5 & 9 & 5$\pm$2 & 0.53 & 10.3$\pm$0.2 
& 12\,000$\pm$2\,000 
\\
\hline
\end{tabular}
\end{center}
\label{table:I}
\end{table}

\subsection{Effects of reduced $\mu$ and increased $R$}

We found $\overline{\tau}$ to be approximately inversely proportional 
to $\mu$ and $R$, while the mean diversity increases (but apparently
sublinearly) with both $\mu$ and $R$. Within the uncertainty, 
$\overline{\tau}$ was the same for the diversity and
population size, and the reported results are therefore based on both
quantities. The mean total population size $\overline{N_{\rm tot}}$
appears to be roughly independent of $\mu$ and approximately linearly
dependent on $R$. This latter proportionality 
(which is also expected from the analytical result, Eq.~(\ref{eq:sol})) 
means that 
$R \mu \propto \overline{N_{\rm tot}} \mu$, 
the average total number of mutant individuals per
generation. Even though the models studied here have strong interspecies
interactions, we note that $\overline{N_{\rm tot}} \mu$
is one half of the {\it fundamental biodiversity
number\/} in Hubbell's neutral theory of biodiversity \cite{HUBB01}. 
This parameter appears to play an analogous role in determining the diversity
in the present models as it does in the neutral theory. 

However, more important than the above results is that none of the observed
power laws changed under parameter variation within this range. In
particular, the exponent for the PDF of individual species
lifetimes remained near $-2$ for all parameter sets, while the
exponent for the  QSS duration PDF remained close to $-1$. In fact,
allowing for the larger overall intensity in the population PSD for
$R=8000$ and to within
the numerical accuracy, both PDFs and PSDs for all three parameter
sets with $L=20$ 
can be overlaid graphically with the ones presented 
for $L=13$, $\mu = 10^{-3}$, and $R=2000$ elsewhere in this paper. 
Based on these trials we therefore conclude that the parameters used in the
main study are also representative for systems with larger $L$ and $R$ and
smaller $\mu$ and are well chosen to study the stationary dynamics of
the models on long time scales. 

Analogous, but less exhaustive, tests for Model A give similar results.
Runs with $N_0 = 16\,000$ and $\mu = 1.25 \times 10^{-4}$ ($N_0 \mu = 2$)
gave $\overline{N_{\rm tot}} \approx 1.5\,N_0$ as in the main simulations
with $N_0 =2000$, and $\overline{D} \approx 3.6$, a statistically
insignificant increase from the value of approximately 3.3 for the
simulations with $N_0 =2000$ and $\mu = 10^{-3}$ \cite{RIKV03}.




\end{document}